\documentstyle[12pt]{article}
\newcommand{\beq}{\begin{equation}}
\newcommand{\eeq}{\end{equation}}
\newcommand{\beqa}{\begin{eqnarray}}
\newcommand{\eeqa}{\end{eqnarray}}
\newcommand{\beqan}{\begin{eqnarray*}}
\newcommand{\eeqan}{\end{eqnarray*}}
\newcommand{\no}{\nonumber}

\newcommand{\ul}{\underline}
\newcommand{\ol}{\overline}
\newcommand{\ra}{\rightarrow}

\newcommand{\ben}{\begin{enumerate}}
\newcommand{\een}{\end{enumerate}}
\newcommand{\bfl}{\begin{flushleft}}
\newcommand{\efl}{\end{flushleft}}
\newcommand{\ba}{\begin{array}}
\newcommand{\ea}{\end{array}}
\newcommand{\btab}{\begin{tabular}}
\newcommand{\etab}{\end{tabular}}
\newcommand{\bit}{\begin{itemize}}
\newcommand{\eit}{\end{itemize}}

\newcommand{\cA}{{\cal A}}

\newcommand{\vs}{\vspace}
\newcommand{\hs}{\hspace}

\def \sam {\triangle}
\newcommand{\prepr}[1] {\begin{flushright} {\bf #1} \end{flushright} \vskip
1.cm}
\newcommand{\titul}[1] {\begin{center}{\Large {\bf #1 } } \end{center}
\vskip 0.8cm}

\newcommand{\autor}[1] {\begin{center} \large {\bf \lineskip .3cm #1  }
                        \end{center} }

\newcommand{\lugar}[1] {\begin{center}  {\normalsize \bf \it #1   }
\end{center}}
\newcommand{\abstr}[1] {{\begin{center} \vskip .5cm {\bf \large Abstract
                        \vspace{0pt}} \end{center}}\begin{quote} \small #1
                        \end{quote}}
\newcounter{muni}

\begin{document}
\vspace{.1cm}
\hbadness=10000
\pagenumbering{arabic}
\begin{titlepage}
{\bf \em 5 January, 1995} \hs{80mm}
\prepr{Preprint PAR/LPTHE/95-01 \hs{4mm} \\ hep-ph/95XXXXXX }
\titul{\large FACTORIZATION AND SU(2) HEAVY FLAVOR SYMMETRY \\
 FOR B-MESON DECAYS PRODUCING CHARMONIUM}
\autor{ M. Gourdin\footnote{\rm Postal address: LPTHE, Tour 16, $1^{er}$ Etage,
4 Place Jussieu, F-75252 Paris CEDEX 05, France.},
Y. Y. Keum\footnote{\rm Postal address: LPTHE, Tour 24, $5^{\grave{e}me}$
Etage,
 2 Place Jussieu, F-75251 Paris CEDEX 05, France. \\
{\small  E-mail : gourdin@lpthe.jussieu.fr, keum@lpthe.jussieu.fr and
pham@lpthe.jussieu.fr.}
}
and X. Y. Pham$^*$ }

\lugar{Universit\'e Pierre {\it \&} Marie Curie, Paris VI \\
Universit\'e Denis Diderot, Paris VII \\
Physique Th\'eorique et Hautes Energies, \\
Unit\'e associ\'ee au CNRS : D 0280
}

\vs{-14cm}
\thispagestyle{empty}
\vs{133mm}
\noindent
\abstr{
\hs{5mm}
We show that the factorization assumption
in color-suppressed $B$ meson decays
is not ruled out by experimental data on
$B \ra K(K^*) + J/\Psi(\Psi^{'})$.
The problem previously pointed out might be
due to an inadequate choice of hadronic form factors.

\hs{5mm}
Within the Isgur-Wise SU(2) heavy flavor symmetry framework,
we search for possible  $q^2$ dependence of
form factors that are capable of explaining simultaneously
the large longitudinal polarization $\rho_L$ observed
in $B \ra K^* + J/\Psi$ and the relatively small ratio of rates
$R_{J/\Psi} = \Gamma(B \ra K^* + J/\Psi)/\Gamma(B \ra K + J/\Psi)$.

\hs{5mm}
We find out that the puzzle could be essentially understood
if the $A_1(q^2)$ form factor is frankly decreasing,
instead of being almost constant or increasing as commonly assumed.

\hs{5mm}
Of course, the possibility of understanding experimental data is
not necessarily a proof of factorization.
}
%

\vs{3mm}
{\bf \small PACS index 13. 25. Hw, 14. 40. Nd }

\end{titlepage}

\newpage
\hspace{1cm} \large{} {\bf I \hs{3mm} Introduction}    \vspace{0.5cm}
\normalsize

In a recent paper \cite{GKP}, Kamal and two of us (M.G. and X.Y.P.) have shown,
within the factorization approach,
the failure of commonly used $B \ra K(K^*)$ form factors
in explaining recent data on $B \ra J/\Psi + K(K^*)$ decays.
The main problem is a simultaneous fit of the large
longitudinal polarization $\rho_L$ in $B \ra J/\Psi + K^*$ decay
and of the  relatively small ratio
$R_{J/\Psi}$ of the $J/\Psi + K^*$ rate compared to the $J/\Psi + K$ one.
We concluded that the difficulty in understanding experimental data
might be due to a failure of the
factorization method or to a wrong choice of the hadronic form factors or both.

Such an analysis has been independently performed by Aleksan, Le Yaouanc,
Oliver, P\`ene and Raynal \cite{Orsay} who also found difficulties
in fitting both $\rho_L$ and $R_{J/\Psi}$,
in spite of their large choice of heavy to light hadronic form factors.

In the previous work \cite{GKP},
in addition to our exploration of the usual $B \ra K(K^*)$ form factors
 available in the literature,
we also related
 the $B \ra K(K^*)$ to the $D \ra K(K^*)$ form factors using the SU(2)
heavy flavor symmetry between the b and c quarks as first proposed
by Isgur and Wise \cite{IW}.
The input data are the hadronic form factors in the $D$ sector
normalized at $q^2 = 0$ as extracted from semi-leptonic
$D \ra \ol{K}(\ol{K^*}) + \ell^+ + \nu_{\ell}$ decay.
In such experiments, the $q^2$ distributions are not measured, and the analysis
of experimental data has been made assuming monopole $q^2$ dependence for all
the $D \ra \ol{K}(\ol{K^*})$ form factors.
For that reason in \cite{GKP}, we have also used monopole forms in the $B$
sector.
The resulting $B \ra K(K^*)$ form factors obtained in this way were also unable
to explain simultaneously $\rho_L$ and $R_{J/\Psi}$.

Our method, based on the Isgur-Wise relations, has been
subsequently adopted by Cheng and Tseng
\cite{Cheng} who considered various types of $q^2$ dependence
for the hadronic form factors.
However their model still encounters difficulties in reproducing correctly
experimental data.

The purpose of this paper is to make a purely phenomenological investigation of
the possible
 $q^2$ dependence - we shall call scenario - of the hadronic form factors in
the $B$ sector
such that, assuming factorization and using the Isgur-Wise relations \cite{IW}
together with the latest data at $q^2 = 0$ in the $D$ sector \cite{RPR},
we are able to obtain a good fit for both $\rho_L$ and $R_{J/\Psi}$.

Some preliminary remarks are in order.
We are aware of the fact that the values at $q^2 = 0$ of
the $D \ra K(K^*)$ form factor
have been extracted from semi-leptonic decay experiments
assuming a monopole $q^2$ dependence
for all hadronic form factors.
This ansatz is certainly inconsistent will  theoretical
expectations coming, for instance, from QCD sum rules \cite{QCDSUM},
from lattice gauge calculations \cite{Lattice}
as well as from asymptotic scaling law of heavy flavours
\cite{{Orsay},{IW},{Cheng}}.
A correct procedure would be to reanalyze
the triple angular distribution fit \cite{RPR}
in the $D$ semi-leptonic decay, with different scenarios,
in order to evaluate  the sensitivity to the scenarios of
the values at $q^2 = 0$ of the form factors.
Such a study has not yet been done by experimentalists.
Of course the cleanest information would come from a measurement of the $q^2$
distributions for the rates and for the various polarizations
in the semi-leptonic $D$ sector.
We are still far from such an ideal situation and for the time being,
the only pragmatic way is to use the results quoted in \cite{RPR}
$\ul{\rm with \hs{2mm} errors \hs{2mm} included}$
for the values at $q^2 =0$ of the form factors.

We propose, in this paper, four types of scenarios for each of the
$B \ra K(K^*)$ hadronic
form factors $F_1, A_1, A_2$ and $V$ in the Bauer, Stech and Wirbel
(BSW henceforth) notation
\cite{BSW}.

The $q^2$ dependences are taken as $(1 - q^2/\Lambda^2)^{-n}$
applied indiscriminately to all of these form factors.

The algebraic integer $n$ symbolically represents
$n_F, n_1, n_2$ and $n_V$ associated respectively to
$F_1, A_1, A_2$ and $V$.
These integers $n$ can take four values corresponding to four types of
scenarios
mentioned above :
$- 1$ for a linear dependence, $0$ for a constant,
$+ 1$ for a monopole and $+ 2$ for a dipole.

The pole masses $ \Lambda_F, \Lambda_1, \Lambda_2 $ and $ \Lambda_V$
for $F_1, A_1, A_2$ and $V$ respectively
are treated as phenomenological parameters.
Being related, in some way, to bound states of the $\ol{b}s$ system,
we impose to these parameters the physical constraint to be
in the range $ (5 - 6)  \hs{2mm}GeV$.
Such a requirement is satisfied by the pole masses of the BSW model
\cite{BSW}.

We now summarize the results of our finding : \\
The experimental data on $\rho_L$ and $R_{J/\Psi}$ indeed
can been fitted for three seenarios corresponding to three possibilities
$n_2 = 2, 1, 0$ for $A_2^{BK^*}$ together with :

\hs{20mm} i) $n_1$ = $- 1$ for a linear decreasing with $q^2$ of $A_1^{BK^*}$

\hs{19mm} ii) $n_V$ = $+ 2$ for a dipole increasing with $q^2$ of $V^{BK^*}$

\hs{18mm} iii) $n_F$ = $+ 1$ for a monopole increasing with $q^2$ of $F_1^{BK}$
\\
For a given selected scenario we have a non empty allowed domain
in the  $ \Lambda_F, \Lambda_1, \Lambda_2, \Lambda_V  $ parameter space.
Therefore we obtain hadronic form factors for $B \ra K (K^*)$
reproducing correctly (within experimental errors)
 $\rho_L$ and $R_{J/\Psi}$ with
the parameters $ \Lambda_F, \Lambda_1, \Lambda_2, \Lambda_V $
physically acceptable.
We now easily understand why previous attempts
\cite{{GKP}, {Orsay}, {Cheng}}  were unsuccessful,
mainly because the decrease with $q^2$ of the form factor
$A_1(q^2)$ has never been seriously considered.
Let us emphasize however that such an unusual $q^2$ behaviour has already
been obtained by Narison \cite{QCDSUM} in the QCD sum rule approach.
Of course our result is not a proof of factorization in the $B$ sector.
It only makes wrong the statement that the failure
in explaining simultaneously
$\rho_L$ and $R_{J/\Psi}$
necessarily implies that factorization breaks down
in this sector.

This paper is organized as follows.
In section II
we give the consequences of factorization for the decay amplitudes
$ B \ra K(K^*) + ( \eta_c, J/\Psi, {\Psi}^{'} ) $
which are color-suppressed processes.
We study the kinematics and we review the available experimental data
for these decay modes.

In section  III
we discuss in some detail the Isgur-Wise relations \cite{IW} and, in
particular,the consistency of scenarios in the $B$ and $D$ sectors
as well as the relations between the parameters
$\Lambda_j (j = F, 1 , 2, V)$ in both sectors.
The case of the form factors $F_0$ and $A_0$,
associated to the spin zero part of the currents, is equally discussed.

Section  IV is devoted to the decay modes $B \ra K(K^*) + \Psi^{'} $,
in which some scenario independent results can be obtained.
The left-right asymmetry $ {\cA}_{LR}^{'}$ between
the two transverse polarizations in
$B \ra K^* + \Psi^{'} $ is found to be large and close to its maximal value.
The fractional longitudinal polarization $\rho^{'}_L$
turns out to be a slowly varying function of
$\Lambda_2$ and the ratio of rates $R_{\Psi^{'}}$
as a function of $\Lambda_2$ and $\Lambda_F$.
The result obtained for $R_{{\Psi}^{'}}$
is consistent with experiment \cite{RPR}.
Our prediction for $\rho_L^{'}$ is compared with
that of Kamal and Sanda \cite{Kamal}
who use seven different scenarios.

Section V is the central part of this paper being related to
the decay modes $B \ra K(K^*) + J/\Psi$. The study of $\rho_L$
and $R_{J/\Psi}$  allows us to select only three surviving scenarios
among the $4^3$ = 64 possible cases
and to constraint the $ \Lambda_F, \Lambda_1, \Lambda_2 , \Lambda_V$ parameter
space.

The comparison between $J/\Psi$ and $\Psi^{'}$
in the final states is studied in section VI
and the results, in the framework of our model, are shown
to be compatible with experiment.
Comparison with the work of Ref.\cite{Kamal} is made in some details.

For the decay modes $B \ra K(K^*) + \eta_c$
where no experimental data are available,
we give in section VII, some predictions for the ratios of rates.

Finally, in section VIII,
we come back to the $D$ sector
in the light of results obtained in the $B$ sector.
Of course in the $B$ and $D$ sectors,
the hadronic form factors $F_1, A_1, A_2, V$
follow the same scenarios - same values of $n_F, n_1, n_2, n_V$
and the poles masses are related via Eqs. (\ref{eq:42}) and (\ref{eq:49})
of section III.
We determine the normalized $q^2$ distributions
for semi-leptonic decays  $D \ra \ol{K} \ell^+ \nu_{\ell}$,
 $D \ra \ol{K^*} \ell^+ \nu_{\ell}$
and for this last mode the integrated longitudinal polarization
$\rho_L^{\it sl}$ and the left-right asymmetry
${\cA}_{LR}^{\it sl}$
between transverse polarizations
 ({\it sl} \hs{2mm}stands for semi-leptonic).

A discussion of the results is given in the conclusion.
A more detailed study of all these topics can be found
in our recent internal report
\cite{InterReport}.

\vs{5mm}
\large{}
{\bf  II. \hs{2mm} Factorization and Kinematics, Experimental Data. }
\vs{0.5cm}
\normalsize

\vs{2mm}
$1^o)$
The two-body decays of the charged and neutral $B$ mesons
discussed in this paper are described,
at the tree level, by the color-suppressed diagram.
Penguin diagrams also contribute to these decays at the one loop level.
However the colorless charmonium states $\ol{c}c$
have to be excited from the vacuum and
for which two or three gluons are needed.
For that reason the penguins are neglected in this paper.

\vs{2mm}
$2^o)$
We consider the decay modes
\beqa
B^{+} &\ra& K^{+}({K^*}^{+}) + \eta_c(J/\Psi, \Psi^{'}) \no \\
& &                                                    \label{eq:1} \\
B^{o} &\ra& K^{o}({K^*}^{o}) + \eta_c(J/\Psi, \Psi^{'}) \no
\eeqa
and we compute the decay amplitudes assuming factorization.
We obtain an expression of the form :
\beq
< \ol{c}c + \ol{s}q | T | \ol{b}q > \hs{5mm} \propto \hs{5mm}
< \ol{c}c | J^{\mu} | 0 > \hs{2mm} < \ol{s}q | J_{\mu} | B >  \label{eq:2}
\eeq
The first term in the right-hand side of Eq.(\ref{eq:2})
involves the  decay constants
$f_{\eta_c}, \hs{1mm} f_{J/\Psi}$ and $f_{\Psi^{'}}$
for $\eta_c, \hs{1mm} J/\Psi$ and $\Psi^{'}$ respectively.
The second term is governed by the hadronic form factors for the $B \ra K$
or $B \ra K^*$ transitions.
As a consequence, the branching ratios have the following structure :
\beq
BR = BR_{0} \hs{2mm} \cdot \hs{2mm}
\left( \frac{f_{\ol{c}c}}{m_B} \right)^2
\hs{2mm} \cdot \hs{2mm} PS
\hs{2mm} \cdot \hs{2mm} FF               \label{eq:3}
\eeq
The common scale $BR_{0}$ contains the Fermi coupling constant $G_{F}$,
the Cabibbo-Kobayashi-Maskawa factors $V_{cb}V_{cs}^*$,
the $B$ meson life time $\tau_B$ , and
the phenomenological BSW constant $a_2$
for color-suppressed processes :
\beq
BR_{0} = \left[ \frac{G_F m_B^2}{\sqrt{2}} \right]^2 \hs{2mm}
|V_{cb}|^2 \hs{2mm} |V_{cs}^*|^2 \hs{2mm}
\frac{m_B}{8 \hs{2mm} \pi} \hs{2mm}
a_2^2 \hs{2mm} \frac{\tau_B}{\hbar}                          \label{eq:4}
\eeq
Being interested only in ratios of decay widths,
we shall not compute the $BR_0$ numerically.

The quantity $PS$ is a dimensionless phase space factor depending only on
 masses of the involved particles.
Because of the small $K^{+}({K^*}^{+}) - K^o({K^*}^{o})$ mass differences,
the numerical values of $PS$ are slightly different for $B^{+}$ and $B^o$
decays.
However these differences  are typically  $O(10^{-3})$,
hence we ignore the mass differences between charged and neutral strange mesons
and the numerical values  given below correspond to $B^{+}$ decays.

The last factor $FF$ depends on the hadronic form factors and it contains
the dynamics of the weak decays.

The results are :
\beqa
&(a)&  B^+ \ra K^+ + \eta_c
\hs{7mm} : \hs{10mm} PS = 0.3265
\hs{5mm}, \hs{10mm} FF = |F_0^{BK}(m_{\eta_c}^2)|^2  \label{eq:5} \\
&(b)&  B^+ \ra K^+ + J/\Psi
\hs{3mm} : \hs{9mm} PS = 0.1296
\hs{5mm}, \hs{10mm} FF = |F_1^{BK}(m_{J/\Psi}^2)|^2  \label{eq:6} \\
&(c)&  B^+ \ra K^+ + \Psi^{'}
\hs{6mm} : \hs{9mm} PS = 0.0575
\hs{5mm}, \hs{10mm} FF = |F_1^{BK}(m_{\Psi^{'}}^2)|^2  \label{eq:7} \\
&(d)&  B^+ \ra {K^*}^+ + \eta_c
\hs{5mm} : \hs{9mm} PS = 0.1218
\hs{5mm}, \hs{10mm} FF = |A_0^{BK^*}(m_{\eta_c}^2)|^2  \label{eq:8} \\
&(e)&  B^+ \ra {K^*}^+ + J/\Psi
\hs{1mm} : \hs{9mm} PS = 0.1399
\hs{5mm}, \hs{10mm}  \no \\
& &  \hs{50mm} FF = |A_1^{BK^*}(m_{J/\Psi}^2)|^2
\hs{2mm} [(a - b x)^2 + 2 \hs{1mm}(1 + c^2 y^2) ]  \label{eq:9} \\
&(f)&  B^+ \ra {K^*}^+ + \Psi^{'}
\hs{4mm} : \hs{8mm} PS = 0.1407
\hs{5mm}, \hs{10mm}  \no \\
& & \hs{50mm} FF = |A_1^{BK^*}(m_{\Psi^{'}}^2)|^2
\hs{2mm} [(a^{'} - b^{'} x^{'})^2 + 2 \hs{1mm}(1 + {c^{'}}^2 {y^{'}}^2)]
\label{eq:10}
\eeqa
The analytic expressions for $a, b, c$ are previously given in Ref.\cite{GKP}
and $a^{'}, b^{'}, c^{'}$ are obtained respectively from $a, b, c$ \hs{1mm}
by the simple substitution $m_{\Psi^{'}}$ to $m_{J/\Psi}$.
We get numerically :
\beqa
&& a = 3.1652 \hs{16mm} b = 1.3084 \hs{16mm} c = 0.4356  \label{eq:11} \\
&& a^{'} = 2.0514 \hs{15mm} b^{'} = 0.5538
\hs{15mm} c^{'} = 0.3092  \label{eq:12}
\eeqa
The ratios of form factors $x, \hs{1mm} y, \hs{1mm} x^{'}, \hs{1mm} y^{'}$ are
defined by :
\beqa
x \equiv x^B(m_{J/\Psi}^2)
= \frac{A_2^{B{K^*}}(m_{J/\Psi}^2)}{A_1^{B{K^*}}(m_{J/\Psi}^2)}
\hs{8mm},
& & \hs{9mm}
y \equiv y^B(m_{J/\Psi}^2)
= \frac{V^{B{K^*}}(m_{J/\Psi}^2)}{A_1^{B{K^*} }(m_{J/\Psi}^2)}
                                                    \label{eq:13} \\
\cr
x^{'} \equiv x^B(m_{\Psi^{'}}^2)
= \frac{A_2^{B{K^*}}(m_{\Psi^{'}}^2)}{A_1^{BK^*}(m_{\Psi^{'}}^2)}
\hs{10mm},
& & \hs{10mm}
y^{'} \equiv y^B(m_{\Psi^{'}}^2)
= \frac{V^{BK^*}(m_{\Psi^{'}}^2)}{A_1^{BK^*}(m_{\Psi^{'}}^2)}
                                                    \label{eq:14}
\eeqa

For each of the $K^* + J/\Psi$ and $K^* + \Psi^{'}$ modes, we have three
possible polarization states,
one is longitudinal (LL) and two are transverse ( $--$, $++$ )
for both final particles.
We now define two interesting quantities :
First, the fractional longitudinal polarization :
\beq
\rho_L = \frac{\Gamma(B \ra K^* + J/\Psi)_{LL}}{\Gamma(B \ra K^* + J/\Psi)}
\hs{10mm}, \hs{10mm}
\rho_L^{'} = \frac{\Gamma(B \ra K^* + \Psi^{'})_{LL}}{\Gamma(B \ra K^* +
\Psi^{'})} \label{eq:15}
\eeq
and second, the left-right asymmetry :
\beqa
&& \cA_{LR} = \frac{\Gamma(B \ra K^* + J/\Psi)_{--} - \Gamma(B \ra K^* +
J/\Psi)_{++}}
{\Gamma(B \ra K^* + J/\Psi)_{--} + \Gamma(B \ra K^* + J/\Psi)_{++}} \hs{20mm}
\no \\
&&
\label{eq:16} \\
&& \cA_{LR}^{'} = \frac{\Gamma(B \ra K^* + \Psi^{'})_{--} - \Gamma(B \ra K^* +
\Psi^{'})_{++}}
{\Gamma(B \ra K^* + \Psi^{'})_{--} + \Gamma(B \ra K^* + \Psi^{'})_{++}}  \no
\eeqa

We get :
\beq
\rho_L = \frac{(a - b x )^2}{(a - b x)^2 + 2 [1 + c^2 \hs{1mm} y^2]}
\hs{10mm}, \hs{10mm}
\rho_L^{'} = \frac{(a^{'} - b^{'} x^{'} )^2}
{(a^{'} - b^{'} x^{'})^2 + 2 [1 + {c^{'}}^2 \hs{1mm} {y^{'}}^2]}  \label{eq:17}
\eeq
\beq
\cA_{LR} = \frac{2 c y }{1 + c^2 \hs{0.5mm} y^2}
\hs{25mm}, \hs{25mm}
\cA_{LR}^{'} = \frac{2 c^{'} y^{'}}
{1 + {c^{'}}^2 {y^{'}}^2}                                         \label{eq:18}
\eeq
We  also introduce four ratios of rates, only three of these ratios are
independent :
\beq
R_{J/\Psi} = \frac{\Gamma(B \ra K^* + J/\Psi)}
{\Gamma(B \ra K + J/\Psi)}
\hs{10mm}, \hs{10mm}
R_{\Psi^{'}} = \frac{\Gamma(B \ra K^* + \Psi^{'})}
{\Gamma(B \ra K + \Psi^{'})}
\label{eq:19}
\eeq
\beq
S = \frac{\Gamma(B \ra K + \Psi^{'})}{\Gamma(B \ra K + J/\Psi)}
\hs{10mm}, \hs{15mm}
S^* = \frac{\Gamma(B \ra K^* + \Psi^{'})}{\Gamma(B \ra K^* + J/\Psi)}
\label{eq:20}
\eeq
Defining two more ratios of form factors :
\beq
z \equiv z^B(m_{J/\Psi}^2) =
\frac{F_1^{BK}(m_{J/\Psi}^2)}{A_1^{BK^*}(m_{J/\Psi}^2)}
\hs{10mm}, \hs{10mm}
z^{'} \equiv z^B(m_{\Psi^{'}}^2) =
\frac{F_1^{BK}(m_{\Psi^{'}}^2)}{A_1^{BK^*}(m_{\Psi^{'}}^2)} \label{eq:21}
\eeq
We obtain :
\beqa
R_{J/\Psi} &=& 1.0793 \hs{2mm}
\frac{(a - b \hs{2mm}x)^2 + 2 ( 1 + c^2 \hs{2mm} y^2)}{z^2} \no \\
          & &                                           \label{eq:22} \\
R_{\Psi^{'}} &=&  2.4455 \hs{2mm}
\frac{ (a^{'} - b^{'} x^{'})^2 + 2 [1 + {c^{'}}^2 {y^{'}}^2 ] }{ {z^{'}}^2 }
\no
\eeqa

For $S$ and $S^*$, we get :
\beq
S = 0.4438 \hs{2mm} \left( \frac{f_{\Psi^{'}}}{f_{J/\Psi}} \right)^2 \hs{2mm}
\left| \frac{F_1^{BK}(m_{\Psi^{'}}^2)}{F_1^{BK}(m_{J/\Psi}^2)} \right|^2
\label{eq:23}
\eeq
\beq
S^* = 1.0057 \hs{2mm} \left( \frac{f_{\Psi^{'}}}{f_{J/\Psi}} \right)^2 \hs{2mm}
\left| \frac{A_1^{BK^*}(m_{\Psi^{'}}^2)}{A_1^{BK^*}(m_{J/\Psi}^2)} \right|^2
\hs{2mm}
\frac{(a^{'} - b^{'}x^{'})^2 + 2 (1 + {c^{'}}^2{y^{'}}^2)}
{(a - b x)^2 + 2 ( 1 + c^2 y^2)} \ \label{eq:24}
\eeq

\vs{2mm}
$3^o)$
The experimental data for decay rates as averaged by PDG \cite{RPR} are given
in Table 1.
We have no experimental information on the  $K\eta_c$ and $K^*\eta_c$ modes.

\vs{3mm}
\begin{table}[thb]
\begin{center}
\begin{tabular}{|c||c||c||}
\hline
Modes
& $B^{+}$
& $B^{o}$  \\
\hline
\hline
$K + \Psi^{'} $ &  $(0.69 \pm 0.31) \hs{1mm} \cdot \hs{1mm} 10^{-3} $
& $ < \hs{2mm} 0.8 \hs{1mm} \cdot \hs{1mm} 10^{-3} $ \\
\hline
$K^* + \Psi^{'} $  &  $ < \hs{2mm} 3.0 \hs{1mm} \cdot \hs{1mm} 10^{-3} $
& $(1.4 \pm 0.9) \hs{1mm} \cdot \hs{1mm} 10^{-3}$  \\
\hline
\hline
$K + J/\Psi$  &  $(1.02 \pm 0.14) \hs{1mm} \cdot \hs{1mm} 10^{-3}$
& $(0.75 \pm 0.21) \hs{1mm} \cdot \hs{1mm} 10^{-3}$  \\
\hline
$K^* + J/\Psi$  &  $(1.7 \pm 0.5) \hs{1mm} \cdot \hs{1mm} 10^{-3}$
& $(1.58 \pm 0.28) \hs{1mm} \cdot \hs{1mm} 10^{-3}$  \\
\hline
\hline
\end{tabular}

\vs{3mm}
Table 1. \\
\end{center}
\end{table}

The ratios $R_{\Psi^{'}}$ and $R_{J/\Psi}$, $S$ and $S^*$ can be estimated
from  data of Table 1 and
the results are shown in Table 2.

\vs{3mm}
\begin{table}[thb]
\begin{center}
\begin{tabular}{|c||c||c||c||}
\hline
Ratio
& $B^{+}$
& $B^{o}$
& $B^{+}, B^o$ combined  \\
\hline
\hline
$R_{\Psi^{'}}$ &   $ < \hs{2mm} 4.35 \pm 1.95 $
& $ > 1.75 \pm 1.12 $
& $ 2.03 \pm 1.59 $ \\
\hline
\hline
$R_{J/\Psi}$  &  $1.67 \pm 0.54$
& $2.11 \pm 0.70$
& $1.83 \pm 0.42$ \\
\hline
\hline
$S$  &  $0.68 \pm 0.32$
& $ < \hs{2mm} 1.07 \pm 0.30$
& $0.68 \pm 0.32$ \\
\hline
\hline
$S^*$  &  $< \hs{2mm} 1.76 \pm 0.52$
& $ 0.89 \pm 0.59$
& $0.89 \pm 0.59$ \\
\hline
\hline
\end{tabular}

\vs{3mm}
Table 2. \\
\end{center}
\end{table}
%
%
A direct measurement of $R_{J/\Psi}$
with both $B^{+}$ and $B^o$
by CLEO II \cite{CLEO II} is consistent
with our estimate given in the last column of Table 2,

\hs{30mm} CLEO II  \hs{20mm} $R_{J/\Psi} = 1.71 \pm 0.34$ \\
In what follows we shall use the constraint $R_{J/\Psi} \leq 2.2$.

\vs{2mm}
$4^o)$
For the $B \ra K^* + J/\Psi$ mode,
we have experimental data for the fractional
longitudinal polarization $\rho_L$ :

\hs{20mm} CLEO II \cite{CLEO II}  \hs{20mm} $\rho_L = 0.80 \pm 0.08 \pm 0.05$

\hs{23mm} CDF \cite{CDF} \hs{24mm} $\rho_L = 0.66 \pm 0.10 ^{+ \hs{2mm}
0.08}_{- \hs{2mm} 0.10} $

\hs{20mm} ARGUS \cite{ARGUS} \hs{21mm} $\rho_L = 0.97 \pm 0.16 \pm 0.15$

Averaging these results with the standard weighted least-squares procedure,
we obtain
\footnote{ A. N. Kamal, private communication } :
$$ \rho_L = 0.780 \pm 0.073 $$
In what follows we shall use the one standard deviation lower limit $\rho_L
\geq 0.7 $ .

We remark that model-independent upper bounds for
$\rho_L$ and $\rho_L^{'}$ can be derived \cite{GKP} :
\beq
\rho_L \leq \frac{a^2}{a^2 + 2} \ = 0.833 \hs{10mm},
\hs{10mm}
\rho_L^{'} \leq \frac{{a^{'}}^2}{{a^{'}}^2 + 2} \ = 0.678 \label{eq:pham}
\eeq
These results are actually the most rigorous consequences of factorization,
their violations imply unquestionably the failure of factorization
hypothesis whatever are the form factors.
For this reason, it will be very interesting
if the errors in the new Argus data \cite{ARGUS}
would be significantly reduced.

\vs{2mm}
$5^o)$
The input data in the $D$ sector are coming from analyses of
the semi-leptonic decays $D \ra \ol{K} + \ell^{+} + \nu_{\ell}$ and
$D \ra \ol{K}^* + \ell^{+} + \nu_{\ell}$.
We shall use the average values for $F_1^{DK}(0)$, $A_1^{DK^*}(0)$,
$x^D(0) \equiv A_2^{DK^*}(0)/A_1^{DK^*}(0)$
and $y^D(0) \equiv V^{DK^*}(0)/A_1^{DK^*}(0)$
as given by the Particle Data Group \cite{RPR}.
These results are shown in Table 3.

\vs{3mm}
\begin{table}[thb]
\begin{center}
\begin{tabular}{|c||c||c||c||}
\hline
$F_1^{DK}(0)$
& $A_1^{DK^*}(0)$
& $x^D(0)$
& $y^D(0)$ \\
\hline
\hline
$0.75 \pm 0.03 $ &
$0.56 \pm 0.04 $ &
$0.73 \pm 0.15 $ &
$1.89 \pm 0.25$ \\
\hline
\hline
\end{tabular}

\vs{3mm}
Table 3. \\
\end{center}
\end{table}
%
The quantities $F_1^{DK}(0)$ and $A_1^{DK^*}(0)$ are determined \cite{RPR}
from semi-leptonic integrated rates.
The ratios $x^D(0)$ and $y^D(0)$ are extracted \cite{RPR}
by fitting the angular distributions in
$D \ra \ol{K}^* + \ell^{+} + \nu_{\ell}$.
The ratio $z^D(0) = F_1^{DK}(0)/A_1^{DK^*}(0)$ is found to be
$z^D(0) = 1.34 \pm 0.11$ from Table 3.

Let us remind that the four values in Table 3 are obtained \cite{RPR}
by assuming monopole $q^2$ behaviour of all form factors,
with pole masses in the $(2.1 - 2.5) \hs{2mm} GeV$ region.

\vs{5mm}
\large{}
{\bf  III. \hs{2mm} The ISGUR-WISE relations due to SU(2) flavor symmetry }
\vs{0.5cm}
\normalsize

$1^o)$
The SU(2) flavor symmetry between the heavy b and c quarks allows
us to derive relations between
the $B \ra K(K^*)$ and $D \ra K(K^*)$
hadronic form factors at the same velocity transfer
though at different momentum transfers.
Calling $t_B$ $(t_D)$ the value of the squared momentum transfer
$q^2$ for $B(D)$ form factors,
we obtain the following kinematic relations :
\beqa
v_b \cdot v_K = v_c \cdot v_K \hs{5mm} &{\rm or}&
\hs{3mm} m_c \hs{1mm} t_B - m_b \hs{1mm} t_D
\hs{1mm} = \hs{1mm} (m_b - m_c)(m_b \hs{2mm} m_c - m_K^2)   \label{eq:25} \\
v_b \cdot v_{K^*} = v_c \cdot v_{K^*} \hs{3mm} &{\rm or}&
\hs{3mm} m_c \hs{1mm} t_B^* - m_b \hs{1mm} t_D^*
\hs{1mm} = \hs{1mm} (m_b - m_c)
(m_b \hs{2mm} m_c - m_{K^*}^2)   \label{eq:26}
\eeqa

In practice, we shall use the experimental data  in the $D$ sector
 at zero momentum transfer
$t_D = t_D^* = 0$.
The corresponding values in the $B$ sector, $t_B^o$ and $t_B^{*o}$
are given from Eqs.(\ref{eq:25}) and (\ref{eq:26}) by
\beqa
 t_B^o &=& ( \frac{m_b}{m_c} \ - 1)(m_b \hs{2mm} m_c - m_K^2) \label{eq:27} \\
 t_B^{*o} &=& ( \frac{m_b}{m_c} \ - 1)
(m_b \hs{2mm} m_c - m_{K^*}^2) \label{eq:28}
\eeqa

The knowledge of the hadronic form factors at $q^2 = 0$ in the $D$ sector
will determine the hadronic form factor values in the $B$ sector
at  $q^2 = t_B^o$ for $B \ra K$ case and
at $q^2 = t_B^{*o}$ for $B \ra K^*$ case.

The values of $t_B^o$ and $t_B^{*o}$ depend on the quark masses $m_b$ and
$m_c$.
We choose, in this paper, $m_b = 4.7$ $GeV$, $m_c = 1.45$ $GeV$ and get
\beqa
t_B^o &=& 14.73 \hs{2mm} GeV^2 \label{eq:29} \\
t_B^{*o} &=& 13.49 \hs{2mm} GeV^2 \label{eq:30}
\eeqa

\vs{2mm}
$2^o)$ We first consider the case of $B \ra K$ and $D \ra K$ form factors.
The matrix elements of the weak current involve two form factors
$f_{+}$ and $f_{-}$ defined by
\beqa
<K|J_{\mu}|D> &=& (p_D + p_K)_{\mu} \hs{2mm} f_{+}^{DK}(q^2)
+ (p_D - p_K)_{\mu} \hs{2mm} f_{-}^{DK}(q^2) \label{eq:31} \\
<K|J_{\mu}|B> &=& (p_B + p_K)_{\mu} \hs{2mm} f_{+}^{BK}(q^2)
+ (p_B - p_K)_{\mu} \hs{2mm} f_{-}^{BK}(q^2) \label{eq:32}
\eeqa
where $q = p_{B,D} - p_K$.

In the BSW basis \cite{BSW}, the spin one and
the spin zero parts of the weak current are separated
and two new form factors $F_1$ and $F_0$ are defined :
\beqa
F_1^{PK}(q^2) &=& f_{+}^{PK}(q^2) \label{eq:33} \\
F_0^{PK}(q^2) &=& f_{+}^{PK}(q^2) + \frac{q^2}{m_P^2 - m_K^2} \ f_{-}^{PK}(q^2)
\label{eq:34}
\eeqa
where $P = B $ or $D$.

The Isgur-Wise relations \cite{IW} are written
\footnote{ The QCD  correction factor
$ [ \frac{\alpha_s(m_b)}{\alpha_s(m_c)} \ ]^{-6/25}$
in the right-hand side
has been omitted in the Isgur-Wise relations
because we are essentially interested,
in this paper,
in ratios of form factors
like $x$, $x^{'}$, $y$, $y^{'}$, $z$ and $z^{'}$.
A numerical estimate of this quantity is 1.15.}
with $f_{+}$ and $f_{-}$ :
\beq
(f_{+} \pm f_{-})^{BK}(t_B) = \left( \frac{m_c}{m_b} \ \right)^{\pm 1/2}
(f_{+} \pm f_{-})^{DK}(t_D) \label{eq:35}
\eeq

The spin one form factor $F_1^{BK}$ is then related to both
$f_{+}^{DK}$ and $f_{-}^{DK}$.
It is convenient to write its expression in the form :
\beq
F_1^{BK}(t_B) = \frac{m_b + m_c}{2 \sqrt{m_b \hs{1mm} m_c}} \
\left[ 1 + \frac{m_b - m_c}{m_b + m_c} \hs{1mm} \mu^D(t_D) \right]
\hs{2mm} F_1^{DK}(t_D) \label{eq:36}
\eeq
where the ratios of two form factors $f_{-}$ and $f_{+}$ are defined by
\beq
\mu^{P}(q^2) = - \hs{2mm} \frac{f_{-}^{PK}(q^2)}{f_{+}^{PK}(q^2)}
\hs{15mm} P = B,D \label{eq:37}
\eeq

The spin zero form factor $F_{0}^{BK}$ can then be written in the form :
\beq
F_0^{BK}(q^2) = [ 1 - \frac{q^2}{m_B^2 - m_K^2} \ \mu^B(q^2)]
\hs{3mm} F_1^{BK}(q^2) \label{eq:38}
\eeq

The two ratios $\mu^B(q^2)$ and $\mu^D(q^2)$ are related by
SU(2) flavor symmetry and from Eq.(\ref{eq:35}), we get
\beq
\mu^B(t_B) = \frac{ (m_b - m_c) + (m_b + m_c) \hs{2mm} \mu^D(t_D)}
{ (m_b + m_c) + (m_b - m_c) \hs{2mm} \mu^D(t_D)} \ \label{eq:39}
\eeq

\vs{2mm}
$3^o)$
As explained previously, we shall use in the $D$ sector
the values of the hadronic form factors at $q^2 = 0$
as coming from semi-leptonic experimental data.
Because of the normalization constraint $F_1^{DK}(0) = F_0^{DK}(0)$
only $f_{+}^{DK}(0)$ is known and we cannot have, in that way,
any information on $f_{-}^{DK}(0)$ and $\mu^D(0)$
\footnote{
In principle, $f_{-}^{DK}(0)$ and $\mu^D(0)$ could be measured in
$D \ra \ol{K} + \mu^{+} + \nu_{\mu}$
namely by looking at polarized muon.} .

We now make an assumption which is natural
and also suggested \cite{QPXU}
in the framework of the SU(2) heavy flavor symmetry.
If $f_{+}^{DK}(q^2)$ and $f_{-}^{DK}(q^2)$ have
the same type of $q^2$ dependence,
no matter how it is,
then the ratio $\mu^D$ is independent of $q^2$ and, using the Isgur-Wise
relations (\ref{eq:35}),
we easily see that the same property extends to the $B$ sector.
In particular, the ratio $\mu^B$ is a constant related to $\mu^D$ by
Eq.(\ref{eq:39}).

Now, if $F_1^{DK}(q^2)$ is written in the form
\beq
F_1^{DK}(q^2) = \frac{F_1^{DK}(0)}{ [1 - \frac{q^2}{\Lambda_{DF}^2}]^{n_F}} \
\label{eq:40}
\eeq
where $n_F$ is some algebraic integer, then, using Eq.(\ref{eq:36})
we obtain a similar expression for $F_1^{BK}(q^2)$ :
\beq
F_1^{BK}(q^2) = \frac{F_1^{BK}(0)}{ [1 - \frac{q^2}{\Lambda_{F}^2}]^{n_F}} \
\label{eq:41}
\eeq
with the same $n_F$.

Furthermore as explained in \cite{{GKP}, {InterReport}},
a relation between the pole masses $\Lambda_{DF}$ and $\Lambda_{F}$
in the $D$ and $B$ sectors can be obtained and already
 given in Ref.\cite{GKP},
the result is :
\beq
m_c \hs{2mm} \Lambda_F^2 - m_b \hs{2mm} \Lambda_{DF}^2 = m_c \hs{2mm} t_B^o
= (m_b - m_c) (m_b \hs{2mm} m_c - m_K^2)                  \label{eq:42}
\eeq
Of course, the values at $q^2 = 0$ of the form factors $F_1^{BK}$
and $F_1^{DK}$ are also
related by the Isgur-Wise relation (\ref{eq:35}).
For details see Ref.\cite{InterReport}.

\vs{2mm}
$4^o)$
An attractive scenario for $F_1^{DK}(q^2)$ suggested
by many theoretical studies
\cite{{Orsay}, {BSW}, {Neubert}}
as well as supported by experimental data \cite{{RPR}, {Witherell}}
is a monopole $q^2$ dependence, $n_F = 1$,
with a pole mass $\Lambda_{DF}$ in the $2$ $GeV$ region.
{}From now on, we make this monopole choice for both $F_1^{DK}$
and $F_1^{BK}$ and we shall restrict
the pole mass in the $B$
sector, $\Lambda_F$, to be in the $(5 - 6) \hs{2mm} GeV$ region
in agreement furthermore with Eq.(\ref{eq:42}).

{}From Eq.(\ref{eq:38}) written now with a constant $\mu^B$,
we obtain for the spin $0$ form factor
$F_0^{BK}$ a different $q^2$ behaviour from that of $F_1^{BK}$.
\beq
F_0^{BK}(q^2) = \frac{
[1 - \frac{q^2}{m_B^2 - m_K^2} \ \mu^B] }
{[ 1 - \frac{q^2}{\Lambda_F^2} ]}  \hs{3mm} F_0^{BK}(0) \label{eq:43}
\eeq
For the particular value of $\mu^B$ :
\beq
\mu^B = \frac{m_B^2 - m_K^2}{\Lambda_F^2}  \label{eq:44}
\eeq
the form factor $F_0^{BK}(q^2)$ becomes independent of $q^2$.
Such a situation has been suggested by some  analyses
\cite{{Cheng}, {QPXU}}
and in this paper $\mu^B$ will be related to $\Lambda_F^2$ by
the equality (\ref{eq:44}).
The ratio $\mu^D$,
 computed from $\mu^B$ using the relation (\ref{eq:39}),
becomes also a function of $\Lambda_F^2$.

\vs{2mm}
$5^o)$
We now study the case of the $B \ra K^*$ and $D \ra K^*$ hadronic form factors.
The matrix elements of the weak current involve four form factors,
$f, g, a_{+}$ and $a_{-}$ in the Isgur-Wise basis \cite{IW}, or $A_1, A_2, V$
and $A_0$
in the BSW basis \cite{BSW}.

For $A_1$ and $V$, the Isgur-Wise relations are very simple \cite{IW}.
Forgetting as previously the QCD factors
$[ \frac{\alpha_s(m_b)}{\alpha_s(m_c)} \ ]^{-6/25}$,
we obtain
\beqa
A_1^{BK^*}(t_B^*) &=&
\sqrt{ \frac{m_b}{m_c} \ } \hs{2mm}
 \frac{m_D + m_{K^*}}{m_B + m_{K^*}} \
\hs{2mm} A_1^{DK^*}(t_D^*) \label{eq:45}  \\
\cr
V^{BK^*}(t_B^*) &=&
\sqrt{ \frac{m_c}{m_b} \ } \hs{2mm}
 \frac{m_B + m_{K^*}}{m_D + m_{K^*}} \
\hs{2mm} V^{DK^*}(t_D^*) \label{eq:46}
\eeqa

As a consequence,
the ratio $y^B$ defined in Eq.(\ref{eq:13})
at $q^2 = t_B^{*o}$ is directly given by $y^D(0)$,
the QCD factor cancels out in this ratio :
\beq
y^B(t_B^{*o}) = \frac{m_c}{m_b} \ \hs{2mm}
\left(  \frac{m_B + m_{K^*}}{m_D + m_{K^*}} \right)^2 \hs{2mm}
y^D(0) \label{eq:47}
\eeq
The dependence with respect to $q^2$ of $A_1$ and $V$
in the $B$ and $D$ sectors is preserved by
the Isgur-Wise relations (\ref{eq:45}) and (\ref{eq:46}).
We choose the forms
\beq
A_1^{BK^*}(q^2) = \frac{A_1^{BK^*}(0)}{ [1 - \frac{q^2}{\Lambda_1^2}]^{n_1}} \
\hs{10mm}, \hs{10mm}
V^{BK^*}(q^2) = \frac{V^{BK^*}(0)}{ [1 - \frac{q^2}{\Lambda_V^2}]^{n_V}} \
\label{eq:48}
\eeq
and analogous expressions in the $D$ sectors with pole masses $\Lambda_{D1}$
and $\Lambda_{DV}$ related to $\Lambda_1$ and $\Lambda_V$
by a formula similar to Eq.(\ref{eq:42}) :
\beq
m_c \hs{2mm} \Lambda_B^2 - m_b \hs{2mm} \Lambda_{D}^2 = m_c \hs{2mm} t_B^{o*}
= (m_b - m_c) (m_b \hs{2mm} m_c - m_{K^*}^2)                  \label{eq:49}
\eeq
where $\Lambda_B = \Lambda_1$ or $\Lambda_V$ and
$\Lambda_D = \Lambda_{D1}$ or $\Lambda_{DV}$.
The algebraic integers $n_1$ and $n_V$ will be restricted to the values
$-1, 0, 1 $ and $2$.

\vs{2mm}
$6^o)$
For the two other form factors, $A_2$ and $A_0$ or $a_{+}$ and $a_{-}$,
the Isgur-Wise relations are similar to the one previously discussed
for the $B \ra K$ and $D \ra K$ form factors, and we have :
\beq
(a_{+} \pm a_{-})^{BK^*}(t_B^{*}) = \left( \frac{m_c}{m_b} \right)^{1 \pm 1/2}
\hs{2mm} (a_{+} \pm a_{-})^{DK^*}(t_D^{*})  \label{eq:50}
\eeq
The form factor $A_2^{BK^*}$ is related to both $a_{+}^{DK^*}$
and $a_{-}^{DK^*}$, and we obtain, neglecting the QCD correction factor :
\beq
A_2^{BK^*}(t_B^{*}) = \frac{1}{2} \hs{1mm}
 \sqrt{ \frac{m_c}{m_b} }
\hs{2mm} \left( 1 + \frac{m_c}{m_b} \right) \hs{2mm}
\frac{m_B + m_{K^*}}{m_D + m_{K^*}} \
\hs{2mm} [ 1 + \frac{m_b - m_c}{m_b + m_c} \hs{2mm} \lambda^D(t_D^{*})]
\hs{2mm} A_2^{DK^*}(t_D^{*})
                                                               \label{eq:51}
\eeq
where the ratios of the form factors $a_{-}$ and $a_{+}$
are defined by
\beq
\lambda^P(q^2) = - \hs{1mm} \frac{a_{-}^{PK^*}(q^2)}{a_{+}^{PK^*}(q^2)} \,
\hs{10mm} P = B, D \label{eq:52}
\eeq
Of course, the ratios $\lambda^B(q^2)$ and $\lambda^D(q^2)$ are related by
SU(2) flavor symmetry and from Eqs.(\ref{eq:50}) and (\ref{eq:52}), we get an
equation similar to Eq.(\ref{eq:39})
\beq
\lambda^B(t_B^{*}) = \frac{(m_b - m_c) + (m_b + m_c) \hs{2mm}
\lambda^D(t_D^{*})}
{(m_b + m_c) + (m_b - m_c) \hs{1mm} \lambda^D(t_D^{*})}  \label{eq:53}
\eeq
Finally, the spin zero axial form factor $A_0^{BK^*}(q^2)$
can be  written with the help of $\lambda^B(q^2)$ :
\beq
A_0^{BK^*}(q^2) = \frac{m_B + m_{K^*}}{2 \hs{2mm} m_{K^*}} \hs{2mm}
A_1^{BK^*}(q^2) -
\frac{m_B - m_{K^*}}{2 \hs{2mm} m_{K^*}} \hs{2mm}
[1 - \frac{q^2}{m_B^2 - m_{K^*}^2} \hs{2mm} \lambda^B(q^2)]
\hs{2mm} A_2^{BK^*}(q^2)                                       \label{eq:54}
\eeq

\vs{2mm}
$7^o)$
We now make an assumption for $a_{+}$ and $a_{-}$ similar to the one
 previously made for $f_{+}$ and $f_{-}$.
If $a_{+}^{DK^*}(q^2)$ and $a_{-}^{DK^*}(q^2)$ have the same type of $q^2$
dependence, no matter how it is,
then the ratio $\lambda^D$ is constant and using the Isgur-Wise
relations (\ref{eq:50}), the same property extends to
the $B$ sector.
 In particular, $\lambda^B$ is also independent of $q^2$ and
related to $\lambda^D$ by Eq.(\ref{eq:53}).

We use for $A_2^{BK^*}(q^2)$ a $q^2$ behaviour of the form :
\beq
A_2^{BK^*}(q^2) = \frac{A_2^{BK^*}(0)}
{ \left[ 1 - \frac{q^2}{\Lambda_2^2} \right]^{n_2} } \   \label{eq:55}
\eeq
and an analogous expression for $A_2^{DK^*}(q^2)$  with a pole mass
$\Lambda_{D2}$
 related to $\Lambda_2$ by Eq.(\ref{eq:49})
in which $\Lambda_B = \Lambda_2$, $\Lambda_D = \Lambda_{D2}$.
The algebraic integer $n_2$ will be restricted to the values
$-1, 0, 1 $ and $2$.

{}From Eq.(\ref{eq:54}) with $\lambda^B$ constant, we see that the form factor
$A_0^{BK^*}(q^2)$ has a somewhat complicated $q^2$ behaviour combining
the one of $A_1^{BK^*}(q^2)$ given in Eq.(\ref{eq:48}) and the one of
$A_2^{BK^*}(q^2)$ given in Eq.(\ref{eq:55}),
the latter being modulated by
a linear factor $( 1 - \frac{q^2 \hs{2mm} \lambda^B}{m_B^2 - m_{K^*}^2} )$.
For the  particular choice of $\lambda^B$, similar to the one previously made
for
$\mu^B$ in Eq.(\ref{eq:44}), i.e :
\beq
\lambda^B = \frac{m_B^2 - m_{K^*}^2}{\Lambda_2^2} \label{eq:56}
\eeq
this linear factor cancels one power in $n_2$ of $A_2^{BK^*}(q^2)$
in Eq.(\ref{eq:54}).

In what follows we shall use the relation (\ref{eq:56}) between
$\lambda^B$ and $\Lambda_2$ and the ratio $\lambda^D$ becomes a function
of $\Lambda_2$ via Eqs.(\ref{eq:53}) and (\ref{eq:56}).

\vs{2mm}
$8^o)$
Consider now the ratio of form factors \hs{1mm} $x^B(q^2)$
defined in Eq.(\ref{eq:13}).
{}From Eqs.(\ref{eq:45}) and (\ref{eq:51}), we get
\beq
x^B(t_B^{*o}) = \frac{1}{2} \hs{2mm} \frac{m_c}{m_b} \hs{2mm}
\left(1 + \frac{m_c}{m_b} \right) \hs{2mm}
\left( \frac{m_B + m_{K^*}}{m_D + m_{K^*}} \right)^2 \hs{2mm}
[1 + \frac{m_b - m_c}{m_b + m_c} \hs{2mm} \lambda^D] \hs{2mm} x^D(0)
\label{eq:57}
\eeq
Because of the presence of the $\lambda^D$ term in Eq.(\ref{eq:57}),
the ratio $x^B(t_B^{*o})$ is a function of $\Lambda_2$.

\vs{2mm}
$9^o)$
The problem is somewhat more complicated for the third ratio $z^B(t_B^{*o})$
defined in Eq.(\ref{eq:21})
which mixes the $B \ra K$ form factor $F_1^{BK}$
with the $B \ra K^*$ form factor $A_1^{BK^*}$.
Isgur-Wise SU(2) flavor symmetry relates $F_1^{BK}(t_B^o)$ to $F_1^{DK}(0)$
and $A_1^{BK^{*}}(t_B^{*o})$ to $A_1^{DK^{*}}(0)$.
Because of the $K^* - K$ mass difference, the quantities $t_B^o$ and $t_B^{*o}$
are different ( See Eqs.(\ref{eq:29}) and (\ref{eq:30}) ), and in order to
estimate $z^B(t_B^{*o})$,
hence $F_1^{BK}$ at $t_B^{*o}$, we need to perform an extrapolation and
as already discussed in subsections 3 and 4, we use a monopole form
with the pole mass $\Lambda_F$ for $F_1^{BK}(q^2)$.

The result is
\beq
F_1^{BK}(t_B^{*o}) = \left(
\frac{ \Lambda_F^2 - t_B^o}{ \Lambda_F^2 - t_B^{*o}} \right)
\hs{2mm} F_1^{BK}(t_B^o) \label{eq:58}
\eeq
and from Eqs.(\ref{eq:36}) and (\ref{eq:45}), we finally obtain :
\beq
z^B(t_B^{*o}) =
\left( \frac{ \Lambda_F^2 - t_B^o}{ \Lambda_F^2 - t_B^{*o}} \right)
\hs{2mm} \frac{1}{2} \hs{2mm}
\left( 1 + \frac{m_c}{m_b} \right) \hs{2mm}
\left( \frac{m_B + m_{K^*}}{m_D + m_{K^*}} \right) \hs{2mm}
\left( 1 + \frac{m_b - m_c}{m_b + m_c} \hs{2mm} \mu^D \right)
\hs{2mm} z^D(0) \label{eq:59}
\eeq

\vs{5mm}
\large{}
{\bf  IV. \hs{2mm} The decay modes ${\bf B \ra K(K^*) + \Psi^{'} }$ }
\vs{0.5cm}
\normalsize

\vs{2mm}
$1^o)$
In the previous section we have computed the ratios of form factors
in the $B$ sector, $x^B(q^2)$, $y^B(q^2)$ and $z^B(q^2)$ at $q^2 = t_B^{*o}$
in terms of their counterparts in the $D$ sector at $q^2 = 0$,
$x^D(0)$, $y^D(0)$ and $z^D(0)$.
The key observation, in this section, is that the value of $t_B^{*o}$
computed in Eq.(\ref{eq:30}) using $m_b = 4.7 \hs{2mm} GeV$ and
$m_c = 1.45 \hs{2mm} GeV$ is remarkably close
\footnote{
Other reasonable choices $^{\cite{RPR}}$ of $m_b$ and $m_c$
(constraint to $m_b - m_c = 3.4 \pm 0.2$ $GeV$ in the HQET scheme)
also lead to similar result : $t_B^{*o}$ is close to $m_{\Psi^{'}}^2$.
 }
to $m_{\Psi^{'}}^2$, such as $\frac{t_B^{*o}}{m_{\Psi^{'}}^2} = 0.9931 $.
It is then justified to neglect the variation of the form factors in the $B$
sector
between $t_B^{*o}$ and $m_{\Psi^{'}}^2$.
Using the Isgur-Wise relation (\ref{eq:47}), we obtain
\beq
y^B(m_{\Psi^{'}}^2) = 1.54 \hs{2mm} y^D(0)  \label{eq:60}
\eeq
and using  the PDG values \cite{RPR}, $y^D(0) = 1.89 \pm 0.25$, we get
\beq
y^{'} = y^B(m_{\Psi^{'}}^2) = 2.91 \pm 0.39  \label{eq:61}
\eeq
The knowledge of $y^{'} = y^B(m_{{\Psi}^{'}}^2)$ determines
the left-right asymmetry $\cA_{LR}^{'}$ previously defined in Eq.(\ref{eq:18}).
The result
\beq
\cA_{LR}^{'} = 0.99 \pm 0.01   \label{eq:62}
\eeq
shows that the dominant transverse amplitude has the helicity $\lambda = -1$,
and in the one standard deviation limit, we predict $\cA_{LR}^{'} > 0.98$.

As a second consequence of the knowledge of $y^{'}$,
we can derive an upper bound for the fractional longitudinal polarization
$\rho_L^{'}$
given in Eq.(\ref{eq:17})
\beq
\rho_L^{'}  \leq \frac{ {a^{'}}^2 }{ {a^{'}}^2 + 2 [1 + {c^{'}}^2
\hs{1mm} {y^{'}}^2 ] } \ \label{eq:63}
\eeq
Using the lower one standard deviation for $y^{'}$, we obtain
\beq
\rho_L^{'} \leq 0.566           \label{eq:64}
\eeq
This upper bound for $\rho_L^{'}$ is significantly smaller
than the theoretical upper bound
( due only to factorization )
$\rho_L^{'} \leq  0.678$ in Eq.(\ref{eq:pham}).
Since $m_{\Psi^{'}}^2 \simeq t^{*o}_B$,
we observe that the results (\ref{eq:62}) and (\ref{eq:64})
are scenario independent and these predictions are direct consequences
of the Isgur-Wise relation (\ref{eq:47}).

\vs{2mm}
$2^o)$
Not only the bound Eq.(\ref{eq:64}), but the quantity
$\rho_L^{'}$ itself can be
computed from $x^{'} = x^B(m_{\Psi^{'}}^2)$
and $y^{'} = y^B(m_{\Psi^{'}}^2)$.
{}From Eq.(\ref{eq:57}), we obtain
\beq
 x^B(m_{\Psi^{'}}^2) =
1.0081 \hs{2mm}[ 1 + 0.5285 \hs{2mm} \lambda^D ] \hs{2mm} x^D(0)  \label{eq:65}
\eeq
and using the PDG value \cite{RPR}, $x^D(0) = 0.73 \pm 0.15$, we get
\beq
 x^B(m_{\Psi^{'}}^2) =
(0.74 \pm 0.15) \hs{2mm}[ 1 + 0.5285 \hs{2mm} \lambda^D ]   \label{eq:66}
\eeq
The fractional polarization $\rho_L^{'}$ depends on $\Lambda_2$ via
the parameter $\lambda^D$ ( See Eqs.(\ref{eq:53}) and (\ref{eq:56}) )
and it turns out to be a slowly increasing function of $\Lambda_2$.
Restricting $\Lambda_2$ to the range $(5 - 6) \hs{2mm} GeV$, we have
\beqa
\rho_L^{'} &=& 0.34 \pm 0.05 \hs{10mm}
{\rm for} \hs{3mm} \Lambda_2 = 5 \hs{2mm} GeV \no \\
& & \label{eq:67} \\
\rho_L^{'} &=& 0.40 \pm 0.04 \hs{10mm}
{\rm for} \hs{3mm} \Lambda_2 = 6 \hs{2mm} GeV \no
\eeqa
The error $\sam \rho_L^{'}$ for $\rho_L^{'}$, combines, in  quadrature,
those of  $x^{'}$ and $y^{'}$ due to the quoted errors in Table 3.

Eq.(\ref{eq:67}) is our prediction for $\rho_L^{'}$,
and let us remark that this quantity is easy to measure.

Estimates of $\rho_L^{'}$ have been previously obtained
by Kamal and Santra\cite{Kamal}.
However their method is different from ours.
They consider seven scenarios  relating $J/\Psi$ to $\Psi^{'}$ modes
and the allowed domains in the $x^{'}, y^{'}$ plane for each scenario are
determined by the experimental constraint on $\rho_L$.
The upper bound for $\rho_L^{'}$ found in \cite{Kamal} is the theoretical upper
bound of Eq.(\ref{eq:pham}) and
the lower bound is slightly scenario-dependent and it varies from $0.48$ to
$0.55$
which is larger than our predictions in Eq.(\ref{eq:67}).

\vs{2mm}
$3^o)$
The ratio $R_{\Psi^{'}}$ defined in Eq.(\ref{eq:19}) can be computed using
$x^{'} = x^B(m_{\Psi^{'}}^2)$,
$y^{'} = y^B(m_{\Psi^{'}}^2)$
and $z^{'} = z^B(m_{\Psi^{'}}^2)$.

{}From the Isgur-Wise relation (\ref{eq:59}), we obtain
\beq
z^B(m_{\Psi^{'}}^2) = 1.4621 \hs{2mm}
\left[ \frac{\Lambda_F^2 - t_B^o}{\Lambda_F^2 - t_B^{*o}} \ \right]
\hs{2mm} [ 1 + 0.5285 \hs{2mm} \mu^D] \hs{2mm}
z^D(0) \label{eq:68}
\eeq
and using the PDG value \cite{RPR}, $z^D(0) = 1.34 \pm 0.11$, we have
\beq
z^B(m_{\Psi^{'}}^2) = (1.96 \pm 0.16) \hs{2mm}
\left[ \frac{\Lambda_F^2 - t_B^o}{\Lambda_F^2 - t_B^{*o}} \ \right]
\hs{2mm} [ 1 + 0.5285 \hs{2mm} \mu^D]  \label{eq:69}
\eeq

The ratio $R_{\Psi^{'}}$ depends on both parameters $\Lambda_2$
(via $x^{'}$ )
and $\Lambda_F$
(via $z^{'}$).
Restricting both pole masses $\Lambda_2$ and $\Lambda_F$ in a range (5 - 6)
$GeV$,
we find two extreme values for $R_{\Psi^{'}}$
\beqa
R_{\Psi^{'}} &=& 1.44 \pm 0.28  \hs{10mm} {\rm for} \hs{10mm}
 \Lambda_F = \Lambda_2 = 5 \hs{2mm} GeV \no \\
& &                                              \label{eq:70} \\
R_{\Psi^{'}} &=&  2.92 \pm 0.54  \hs{10mm} {\rm for} \hs{10mm}
 \Lambda_F = \Lambda_2 = 6 \hs{2mm} GeV \no
\eeqa
The experimental estimate  in Table 2 is $2.03 \pm 1.59$ and
due to the large experimental error,
our results in Eq.(\ref{eq:70}) are compatible with experiment
for all values of $\Lambda_2$ and $\Lambda_F$
in the $(5 - 6) \hs{2mm} GeV$ range.

\vs{5mm}
\large{}
{\bf V. \hs{2mm} The decay modes $ B \ra K(K^*) + J/\Psi $ }
\vs{0.5cm}
\normalsize

\vs{2mm}
$1^o)$
In order to compute the longitudinal polarization $\rho_L$
defined in Eq.(\ref{eq:15})
for the $B \ra K^* + J/\Psi$ decay mode and the ratio of the $K^* + J/\Psi$
over $K + J/\Psi$ rates, $R_{J/\Psi}$, defined in Eq.(\ref{eq:19}),
we must extrapolate the ratios of hadronic form factors, $x^B(q^2)$, $y^B(q^2)$
and $z^B(q^2)$ from the value $q^2 = t_B^{*o}$
( where these quantities are given
by the Isgur-Wise relations (\ref{eq:47}), (\ref{eq:57})
and (\ref{eq:59}) ) to the value $q^2 = m_{J/\Psi}^2$.

Of course such an extrapolation is scenario dependent.
We use the pole type $q^2$ dependence as given in Eqs.(\ref{eq:41}),
(\ref{eq:48})
and (\ref{eq:55}), and we introduce the function $r(\Lambda)$ defined by
\beq
r(\Lambda) = \frac{\Lambda^2 - t_B^{*o}}{\Lambda^2 - m_{J/\Psi}^2}
\label{eq:71}
\eeq
and we obtain
\beqa
x  \equiv x^B(m_{J/\Psi}^2) &=&
\frac{[r(\Lambda_2)]^{n_2}}{[r(\Lambda_1)]^{n_1}}
\hs{2mm} x^B(t_B^{*o})            \label{eq:72}  \\
y  \equiv y^B(m_{J/\Psi}^2) &=&
\frac{[r(\Lambda_V)]^{n_V}}{[r(\Lambda_1)]^{n_1}}
\hs{2mm} y^B(t_B^{*o})            \label{eq:73}  \\
z  \equiv z^B(m_{J/\Psi}^2) &=&
\frac{[r(\Lambda_F)]^{n_F}}{[r(\Lambda_1)]^{n_1}}
\hs{2mm} z^B(t_B^{*o})            \label{eq:74}
\eeqa
In our model $n_F = 1$, and for the three other powers $n_1, n_2, n_V$,
each one can  take four algebraic integers $-1, 0, 1, 2$.
On physical grounds, we impose to the pole masses $\Lambda_1, \Lambda_2,
\Lambda_V,
\Lambda_F$
to be inside the (5 - 6) $GeV$ range
where the $\ol{b}s$ bound states masses are expected to be.

\vs{2mm}
$2^o)$
We first study the scenario constraints due to  $\rho_L$
and for the $4^3 = 64$ possible triplets $[n_1, n_2, n_V]$,
we have computed $\rho_L$ using the values of $x$ and $y$
as given  from $x^{'}$ and $y^{'}$ by Eqs.(\ref{eq:61}), (\ref{eq:66}),
(\ref{eq:72}) and (\ref{eq:73}).
We impose the experimental constraint in the form
$\rho_L + \sam \rho_L \geq 0.70 $
where the theoretical error $\sam \rho_L$ is computed in quadrature from
those of $x^D(0)$ and $y^D(0)$.

After a long numerical scanning,
our results can be summarized as follows :

\hs{10mm} i) \hs{2mm}
no solution is obtained when $n_1 = 0, 1, 2$ for all the 16 values of the
couple
$[n_2, n_V]$.

\hs{10mm} ii) \hs{2mm}
solutions exist only when $n_1 = -1$, i.e. when the hadronic form factor
$A_1^{BK^*}(q^2)$ exhibits
a linear decrease with $q^2$.
Of course, in this case, $\Lambda_1$ is no more a pole mass but only a slope
coefficient
and it is reasonable now to relax the constraint $\Lambda_1 \leq 6 \hs{2mm}
GeV$ and
to use only $\Lambda_1 \geq 5 \hs{2mm} GeV$ in order to exclude a too fast
variation with
$q^2$ of $A_1^{BK^*}(q^2)$.

\hs{10mm} iii) \hs{2mm}
The solutions obtained correspond to only four triplets $[n_1, n_2, n_V]$ :
$$
[-1, 2, 2], \hs{3mm} [-1, 1, 2], \hs{3mm} [-1, 0, 2], \hs{3mm} [-1, 2, 1]
$$
and in the four cases,
the maximal value of $\rho_L$
occurs at $\Lambda_1 = 5 \hs{2mm} GeV$, $\Lambda_2 = 6 \hs{2mm} GeV$,
$\Lambda_V = 5 \hs{2mm} GeV$  and in the most favorable situation of two dipole
$q^2$ dependence for $A_2$ and $V$,
we obtain $\rho_L = 0.7162 \pm 0.0236$.
Therefore $\rho_L = 0.74$ is the maximal value within one standard deviation
we can produce in our approach, considering only the quantity $\rho_L$.

\vs{2mm}
$3^o)$
We now consider the ratio of rates $R_{J/\Psi}$ and
we impose the experimental constraint in the form
$R_{J/\Psi} - \sam R_{J/\Psi} \leq 2.2$
where the theoretical error $\sam R_{J/\Psi}$ is computed in quadrature from
the
experimental errors on $x^D(0)$, $y^D(0)$ and $z^D(0)$.

On the one hand, the constraint $\rho_L + \sam \rho_L \geq 0.7$ has selected
four scenarios
previously discussed
and at fixed $\Lambda_2, \Lambda_V$
the allowed domain for $\Lambda_1$ is defined by
an upper limit for $\Lambda_1$ :
\beq
\Lambda_1 \leq \Lambda_{1,\hs{1mm}MAX}(\Lambda_2, \Lambda_V) \label{eq:75}
\eeq

On the other hand, the constraint  $R_{J/\Psi} - \sam R_{J/\Psi} \leq 2.2$
implies a lower limit for $\Lambda_1$ :
\beq
\Lambda_1 \geq \Lambda_{1,\hs{1mm}MIN}(\Lambda_2, \Lambda_V, \Lambda_F)
\label{eq:76}
\eeq
Acceptable values of $\Lambda_1$ exist when and only when
the lower limit (\ref{eq:76})
is smaller than the upper limit (\ref{eq:75}).

The quantity $\Lambda_{1, \hs{1mm}MIN}$ ( at fixed $\Lambda_2, \Lambda_V$ )
is an increasing function of $\Lambda_F$ and using the constraint
$\Lambda_F \geq 5 \hs{2mm} GeV$,
the physical domain for $\Lambda_1$,
at fixed $\Lambda_2, \Lambda_V$, is defined by
\beq
\Lambda_{1, \hs{1mm}MIN}(\Lambda_2, \Lambda_V, \Lambda_F = 5 \hs{2mm} GeV)
\hs{1mm}
\leq \hs{1mm} \Lambda_1 \hs{1mm} \leq \hs{1mm}
\Lambda_{1, \hs{1mm}MAX}(\Lambda_2, \Lambda_V)   \label{eq:77}
\eeq

We find out that for the scenario $[n_1, n_2, n_V] = [-1, 2, 1]$,
the inequality (\ref{eq:77}) has no solution.
For the three remaining scenarios
$[n_1, n_2, n_V] = [-1, 2, 2], [-1, 1, 2]$ and $[-1, 0, 2]$,
the physical domains in the $\Lambda_1, \Lambda_2,\Lambda_V$ space
are represented on Figures 1, 2 and 3.

At fixed $\Lambda_2, \Lambda_V$, we also have
\beq
5 \hs{2mm} GeV \leq \Lambda_F(\Lambda_2, \Lambda_V)
\leq \Lambda_{F, \hs{1mm}MAX}(\Lambda_2, \Lambda_V) \label{eq:78}
\eeq
where $\Lambda_{F,\hs{1mm}MAX}(\Lambda_2, \Lambda_V)$ is determined by
\beq
\Lambda_{1, \hs{1mm}MIN} [ \Lambda_2, \Lambda_V, \Lambda_{F,
\hs{1mm}MAX}(\Lambda_2, \Lambda_V) ]
= \Lambda_{1, \hs{1mm}MAX}(\Lambda_2, \Lambda_V)       \label{eq:79}
\eeq
The quantity $\Lambda_{F, \hs{1mm}MAX}(\Lambda_2, \Lambda_V)$
in the $\Lambda_2, \Lambda_V,\Lambda_F$ space
has been represented
on Figures 4, 5 and 6
for the three surviving scenarios respectively.

\vs{2mm}
$4^o)$
Starting with 64 scenarios for the  $q^2$ dependence of the hadronic form
factors
$A_1^{BK^*}, A_2^{BK^*}$ and $V^{BK^*}$,
we have obtained three surviving scenarios,
$n_1 = -1, n_V = 2$ and $n_2 = 2, 1, 0$
for which it is possible to satisfy simultaneously both experimental
constraints :
$\rho_L + \sam \rho_L \geq 0.70$  which produces
an upper bound on
$\Lambda_1$, and
$R_{J/\Psi} - \sam R_{J/\Psi} \leq 2.2$
which implies a lower bound on $\Lambda_1$.

In order to have a more precise feeling concerning the nature of the fit
 obtained with our model,
we compute $\rho_L$ and $R_{J/\Psi}$ for values of
$\Lambda_1, \Lambda_2, \Lambda_V $ and $\Lambda_F$
belonging to the physical domains of every scenario
represented in Figures $1 - 6$.
For illustration,
we take some values of
$\Lambda_j$ : $\Lambda_2 = 6 \hs{2mm} GeV$,
$\Lambda_V = \Lambda_F = 5 \hs{2mm} GeV$, and for $\Lambda_1$,
the two extreme values, $\Lambda_{1, \hs{1mm}MAX}$ and $\Lambda_{1,
\hs{1mm}MIN}$.
Of course for $\Lambda_1 = \Lambda_{1, \hs{1mm}MAX}$, we get
$\rho_L + \sam \rho_L = 0.70$ and
for $\Lambda_1 = \Lambda_{1, \hs{1mm}MIN}$, we have
$R_{J/\Psi} - \sam R_{J/\Psi} = 2.2$.
The results are shown on  Table 4
where the numerical values of $\Lambda_{1, \hs{1mm}MAX}$
and $\Lambda_{1, \hs{1mm}MIN}$ are given.

A first observation coming from Table 4 is to realize how difficult
it is to fit simultaneously the large experimental value of $\rho_L$
and the relatively small experimental value of $R_{J/\Psi}$.
The opposite trend between $\rho_L$ and $R_{J/\Psi}$ making the fit
so difficult has been also noticed \cite{Orsay}.
The theoretical relative error coming from $R_{J/\Psi}$ is larger than
the one coming from  $\rho_L$,
and this feature is welcome for obtaining a two-fold fit.
It is essentially due to the fact that $R_{J/\Psi}$, in addition to
the errors on $x^D(0)$ and $y^D(0)$ (as for $\rho_L$),
has a third source of uncertainty due to the errors on $z^D(0)$.
While the theoretical  relative error on $\rho_L$ is
only  between $4 \%$ and $6 \%$,
the one on $R_{J/\Psi}$ is between $ 18 \%$ and $ 24\%$.

A second observation, coming from both Figure 1 and Table 4,
is that the scenario with
a dipole form factor $A_2$, $n_2 = 2$,
is the one with the largest domain in the
$\Lambda_1, \Lambda_2, \Lambda_V $ and $\Lambda_F$ space.
Therefore in this scenario it is relatively easy to accommodate both $\rho_L$
and $R_{J/\Psi}$.
{}From Table 4, the largest possible value of $\rho_L$
we can obtain, in the one standard deviation limit, is
$\rho_L + \sam \rho_L = 0.722$ and the smallest possible value of
$R_{J/\Psi}$, in the one standard deviation limit, is
 $R_{J/\Psi} - \sam R_{J/\Psi} = 1.581$.
These extreme values of $\rho_L$ and $R_{J/\Psi}$
in our model do not occur simultaneously
but at the two different extreme values of $\Lambda_1$.
\vs{3mm}
\begin{table}[thb]
\begin{center}
\begin{tabular}{||c||c|||c||}
\hline
$\Lambda_1(GeV)$ &
$\rho_L$  &
$ R_{J/\Psi} $ \\
\hline
\hline
\multicolumn{3}{||c||}{$n_2 = 2$ } \\
\hline
$\Lambda_{1, \hs{1mm}MAX} = 8.112 $ &
$ 0.665 \pm 0.035 $ &
$2.089 \pm 0.508$  \\
\hline
$\Lambda_{1, \hs{1mm}MIN} = 5.426 $ &
$ 0.694 \pm 0.028 $ &
$2.694 \pm 0.494$ \\
\hline
\hline
\multicolumn{3}{||c||}{$n_2 = 1$ } \\
\hline
$\Lambda_{1, \hs{1mm}MAX} = 6.113 $ &
$ 0.663 \pm 0.037 $ &
$2.324 \pm 0.524$  \\
\hline
$\Lambda_{1, \hs{1mm}MIN} = 5.426 $ &
$ 0.681 \pm 0.033 $ &
$2.714 \pm 0.514$ \\
\hline
\hline
\multicolumn{3}{||c||}{$n_2 = 0$ } \\
\hline
$\Lambda_{1, \hs{1mm}MAX} = 5.292 $ &
$ 0.660 \pm 0.040 $ &
$2.625 \pm 0.542$  \\
\hline
$\Lambda_{1, \hs{1mm}MIN} = 5.183 $ &
$ 0.665 \pm 0.038 $ &
$2.739 \pm 0.539$ \\
\hline
\hline
\end{tabular}

\vs{3mm}
Table 4 \\
\end{center}
\end{table}

\vs{2mm}
$5^o)$
The left-right asymmetry $\cA_{LR}$ defined in Eq.(\ref{eq:18})
has not been experimentally measured.
It can be easily computed in our model.
Depending only on the ratio $y = y^B(m_{J/\Psi}^2)$,
we use Eq.(\ref{eq:73}) with $n_1 = -1$ and $n_V = 2$
and we vary the parameters
$\Lambda_1$ and $\Lambda_V$ inside the allowed domains
represented on Figures 1, 2, and 3 for the three scenarios
$n_2 = 2, 1, 0$.

The results are :
\beqa
(i) \hs{5mm} n_2 = 2 & & \hs{20mm} 0.867 < \hs{2mm} \cA_{LR}  \hs{2mm} < 0.945
\label{eq:102} \\
(ii) \hs{4mm} n_2 = 1 & & \hs{20mm} 0.837 < \hs{2mm} \cA_{LR}  \hs{2mm} < 0.910
\label{eq:103} \\
(iii) \hs{3mm} n_2 = 0 & & \hs{20mm} 0.837 < \hs{2mm} \cA_{LR}  \hs{2mm} <
0.856 \label{eq:104}
\eeqa
The left-right asymmetry in the decay mode $B \ra K^* + J/\Psi$
is large in the three selected scenarios, not as large
as that of the decay mode  $B \ra K^* + \Psi^{'}$ where it has been predicted
to be close to unity ( Eq.(\ref{eq:62}) ).
We observe that the differences in the predictions of the three scenarios
are moderate.

\vs{5mm}
\large{}
{\bf VI. \hs{2mm} Comparison of $J/\Psi$ and $\Psi^{'}$ production }
\vs{0.5cm}
\normalsize

\vs{2mm}
$1^o)$
The ratios of decay widths $S$ and $S^*$ defined in Eq.(\ref{eq:20})
involve the same strange meson, $K$ or $K^*$,
and two different charmonium states
$\Psi^{'}$ and $J/\Psi$,
hence two different leptonic decay constants
$f_{\Psi^{'}}$
and $f_{J/\Psi}$ are involved.
Using \cite{Neubert}
$f_{J/\Psi} = (384 \pm 14) \hs{2mm} MeV$ and
$f_{\Psi^{'}} = (282 \pm 14) \hs{2mm} MeV$
as estimated from the decays $J/\Psi \ra e^{+}e^{-}$ and $\Psi^{'} \ra
e^{+}e^{-}$,
we obtain :
\beq
\left( \frac{f_{\Psi^{'}}}{f_{J/\Psi}} \right)^2 = 0.539 \pm 0.066
\label{eq:83}
\eeq
and the quantities $S$ and $S^*$ are written from Eqs.(\ref{eq:23}) and
(\ref{eq:24})
in the form :
\beq
S = [0.2392 \pm 0.0292] \hs{2mm}
\left| \frac{F_1^{BK}(m_{\Psi^{'}}^2)}{F_1^{BK}(m_{J/\Psi}^2)} \right|^2
\label{eq:84}
\eeq
\beq
S^* = [0.5421 \pm 0.0664] \hs{2mm}
\left| \frac{A_1^{BK^*}(m_{\Psi^{'}}^2)}{A_1^{BK^*}(m_{J/\Psi}^2)} \right|^2
\hs{2mm}
\frac{(a^{'} - b^{'}x^{'})^2 + 2 (1 + {c^{'}}^2{y^{'}}^2)}
{(a - b x)^2 + 2 ( 1 + c^2 y^2)} \ \label{eq:85}
\eeq

\vs{2mm}
$2^o)$
In our model the hadronic form factor $F_1^{BK}(q^2)$ has a monopole $q^2$
dependence
with a pole mass $\Lambda_F$ and we simply have :
\beq
\frac{F_1^{BK}(m_{\Psi^{'}}^2)}{F_1^{BK}(m_{J/\Psi}^2)}
= \frac{\Lambda_F^2 - m_{J/\Psi}^2}{\Lambda_F^2 - m_{\Psi^{'}}^2} \label{eq:86}
\eeq
This ratio of form factor is a decreasing function of $\Lambda_F^2$
and so is the ratio $S$.
At $\Lambda_F = 5 \hs{2mm} GeV$, we obtain :
\beq
S(\Lambda_F = 5 \hs{2mm} GeV) = 0.4363 \pm 0.0537 \label{eq:87}
\eeq
This prediction is in agreement, within one standard deviation,
 with the experimental value
estimated in Table 2, $S_{exp} = 0.68 \pm 0.32$.
Such an agreement continues to occur for larger values of $\Lambda_F$ up to
$6.27$ $GeV$.

The range of $\Lambda_F$ depends on the three scenarios corresponding to $n_2 =
2, 1, 0$
and they are deduced from Figures 4, 5, 6 respectively.
We get :
\beqa
\hs{3mm}(i) \hs{2mm}
n_2 = 2 :\hs{7mm} & &
0.3505 \pm 0.0432
\hs{2mm} \leq \hs{2mm}
S \hs{2mm} \leq \hs{2mm}
0.4363 \pm 0.0537      \label{eq:88} \\
\hs{2mm}(ii) \hs{2mm}
n_2 = 1 : \hs{7mm} & &
0.3790 \pm 0.0467
\hs{2mm} \leq \hs{2mm}
S \hs{2mm} \leq \hs{2mm}
0.4363 \pm 0.0537       \label{eq:89} \\
\hs{1mm}(iii) \hs{2mm}
n_2 = 0 : \hs{7mm} & &
0.4181 \pm 0.0515
\hs{2mm} \leq \hs{2mm}
S \hs{2mm} \leq \hs{2mm}
0.4363 \pm 0.0537       \label{eq:90}
\eeqa
The errors quoted in Eq.(\ref{eq:88}), (\ref{eq:89}) and (\ref{eq:90})
are due to the uncertainty on the leptonic decay constants
$f_{\Psi^{'}}$ and $f_{J/\Psi}$.
In conclusion, the theoretical predictions of our model
for the three scenarios
agree, within one standard deviation, with experimental results.

\vs{2mm}
$3^o)$
The analysis of the second ratio $S^*$ is more complex because of
a large number of hadronic form factors involved.
In our model the form factor $A_1^{BK}(q^2)$ has a decreasing linear
$q^2$ dependence with a pole mass $\Lambda_1$, and we simply have
\beq
\frac{A_1^{BK^*}(m_{\Psi^{'}}^2)}{A_1^{BK^*}(m_{J\Psi}^2)}
= \frac{\Lambda_1^2 - m_{\Psi^{'}}^2}{\Lambda_1^2 - m_{J/\Psi}^2} \label{eq:91}
\eeq
We have computed the ratio $S^*$ for the three scenarios $n_2 = 2, 1, 0$
using the allowed domains represented respectively on Figures 1, 2 and 3
for $\Lambda_1, \Lambda_2$ and $\Lambda_V$.

The results of this scanning are :
\beqa
\hs{3mm} (i) \hs{2mm}
n_2 = 2 : \hs{10mm} & & 0.3287 \pm 0.0028 \hs{2mm} \leq \hs{2mm}
S^* \hs{2mm} \leq \hs{2mm} 0.4135 \pm 0.0038   \label{eq:92}  \\
\hs{2mm} (ii) \hs{2mm}
n_2 = 1 : \hs{10mm} & & 0.3489 \pm 0.0034 \hs{2mm} \leq \hs{2mm}
S^* \hs{2mm} \leq \hs{2mm} 0.4015 \pm 0.0039   \label{eq:93}  \\
\hs{1mm} (iii) \hs{2mm}
n_2 = 0 : \hs{10mm} & & 0.3763 \pm 0.0039 \hs{2mm} \leq \hs{2mm}
S^* \hs{2mm} \leq \hs{2mm} 0.3867 \pm 0.0040   \label{eq:94}
\eeqa
The errors quoted in Eqs.(\ref{eq:92}), (\ref{eq:93}) and (\ref{eq:94}) are
computed in  quadrature from those on the ratios
$f_{\Psi^{'}}/f_{J/\Psi}$, $x^D(0)$ and $y^D(0)$.
The theoretical predictions of our model for the three scenarios agree,
 within one standard deviation,
 with the experimental results estimated in Table 2 :
$S_{exp}^* = 0.89 \pm 0.59$.

\vs{2mm}
$4^o)$
Kamal and Santra \cite{Kamal} have studied the ratios $S$ and $S^*$
denoted by them respectively as $1/R$ and $1/R^{'}$.
In the case of $R$, both monopole and dipole $q^2$ dependences for $F_1^{BK}$
are considered with a pole mass $\Lambda_F = 5.43 \hs{2mm} GeV$.
 Their conclusion is that a dipole behaviour for $F_1^{BK}$ is
needed in order to obtain for $R$ agreement between theory
and experiment in the one standard deviation limit.

The apparent contradiction between our result
( monopole for $F_1^{BK}$ ) and the one of Ref.\cite{Kamal}
is essentially due to the large experimental error of $47 \%$ for the quantity
$S$ or $R$.
With $\delta = 0.47$ the relation at first order in $\delta$,
$(1 \pm \delta)^{-1} = 1 \mp \delta$ is not valid and one standard deviation
limit
for $S$ and one standard deviation limit for $R$ are different concepts.
However, since the main part of the experimental error is due to the $K +
\Psi^{'}$
mode and for that reason the consideration of one standard deviation for $S$
( where $K + \Psi^{'}$ enters in the numerator )
seems to be more relevant than for $R$.

A similar situation occurs for $S^*$ and $R^{'}$.
Here the experimental error is even larger, $66.7 \%$, and it is mainly due to
the $K + \Psi^{'}$ mode which enters in the numerator of $S^*$.
Again the one standard deviation limit for $S^*$ and
the one standard deviation limit for $R^{'}$ are different quantities.

Also the pole masses in Ref.\cite{Kamal} are taken only at some fixed values,
while in our approach these poles sweep inside the $(5 - 6) \hs{2mm} GeV$
range.

Furthermore,
considering only the one standard deviation limit for $R^{'}$,
they exclude four scenarios where $A_1^{BK^*}$ is either constant
or linearly decreasing with $q^2$ and conclude that
  if factorization assumption were to hold, then the only scenarios that
are consistent with experiment are those in which $A_1^{BK^*}$ rises with
$q^2$.We observe however that $R^{'}$ (or $S^*$) is not an independent ratio
but related to the other ratios by $S^* R_{J/\Psi} = S R_{{\Psi}^{'}}$,
such that considering  $R^{'}$ (or $S^*$) alone might be inadequate.

\vs{5mm}
\large{}
{\bf VII. \hs{2mm} The decay modes $B \ra K(K^*) + \eta_c$ }
\vs{0.5cm}
\normalsize

\vs{2mm}
$1^o)$
The decay modes $B \ra K + \eta_c$ and $B \ra K^* + \eta_c$ have not been
experimentally observed.
However their rates can be easily computed  and
the relevant expressions have been given in Eqs.(\ref{eq:5}) and (\ref{eq:8}).
The form factors involved in these modes correspond to
the spin zero part of the weak current, $F_0^{BK}(m_{\eta_c}^2)$ and
$A_0^{BK^*}(m_{\eta_c}^2)$.

\vs{2mm}
$2^o)$
The hadronic form factor $F_0^{BK}(t_B^o)$ can be computed using the Isgur-Wise
relations (\ref{eq:36}) together with Eqs.(\ref{eq:38}) and  (\ref{eq:39}).
The result is
\beq
F_0^{BK}(t_B^o) = \frac{m_b + m_c}{2 \hs{1mm} \sqrt{m_b \hs{1mm} m_c} }
\hs{2mm}
\left( \left[ 1 - \frac{m_b - m_c}{m_b + m_c} \hs{2mm} \frac{t_B^o}{m_B^2 -
m_K^2} \right]
- \left[ \frac{t_B^o}{m_B^2 - m_K^2} - \frac{m_b - m_c}{m_b + m_c}  \ \right]
\hs{2mm}
\mu^D   \right) \hs{2mm} F_1^{DK}(0)      \label{eq:95}
\eeq
Numerically we obtain
\beq
F_0^{BK}(t_B^o) = 0.8460 \hs{2mm} [ 1 - 0.00667 \hs{2mm} \mu^D] \hs{2mm}
F_1^{DK}(0)
\label{eq:96}
\eeq
We notice that the coefficient of $\mu^D$ is very small
in the bracket of Eq.(\ref{eq:96}) and
as a consequence, $F_0^{BK}(t_B^o)$ depends only weakly on $\mu^D$.

In our model $F_0^{BK}$ is constant and therefore Eq.(\ref{eq:96})
gives its value for all $q^2$.
The weak dependence on $\mu^D$ implies a weak dependence of $F_0^{BK}$
on the pole mass $\Lambda_F$.

\vs{2mm}
$3^o)$
The hadronic form factor $A_0^{BK^*}$ is deduced from $A_1^{BK^*}$ and
$A_2^{BK^*}$
by using Eq.(\ref{eq:54}) with $\lambda^B$ fixed by the relation (\ref{eq:56}).
For $q^2 = m_{\eta_c}^2$, we have
\beq
A_0^{BK^*}(m_{\eta_c}^2) = \frac{m_B + m_{K^*}}{2 \hs{2mm} m_{K^*}}
\hs{2mm} A_1^{BK^*}(m_{\eta_c}^2)
- \frac{m_B - m_{K^*}}{2 \hs{2mm} m_{K^*}} \hs{2mm} \left[ 1 -
\frac{m_{\eta_c}^2}{\Lambda_2^2} \right]
\hs{2mm} A_2^{BK^*}(m_{\eta_c}^2)      \label{eq:97}
\eeq
In order to obtain $A_1^{BK^*}(m_{\eta_c}^2)$ and $A_2^{BK^*}(m_{\eta_c}^2)$,
we extrapole these form factors
$A_1^{BK^*}$ and $A_2^{BK^*}$ from the value $q^2 = t_B^{*o}$
$-$ where these terms are given by the Isgur-Wise relations
(\ref{eq:45}) and (\ref{eq:51}) $-$
to $q^2 = m_{\eta_c}^2$.
The form factor $A_0^{BK^*}(m_{\eta_c}^2)$ is scenario dependent,
firstly on  $n_2$,
secondly on $\Lambda_1$ and $\Lambda_2$ restricted
to the allowed domains of Figures 1, 2 and 3.

\vs{2mm}
$4^o)$
To bypass the unknown decay constant $f_{\eta_c}$, we consider
the ratio of rates $R_{\eta_c}$ defined by
\beq
R_{\eta_c} = \frac{\Gamma(B \ra K^* + \eta_c)}
{\Gamma(B \ra K + \eta_c)} \label{eq:98}
\eeq
This quantity is given from Eq.(\ref{eq:5}) and (\ref{eq:8}) by :
\beq
R_{\eta_c} = 0.3732 \hs{2mm}
\left|
\frac{A_0^{BK^*}(m_{\eta_c}^2)}{F_0^{BK}(m_{\eta_c}^2)}
\right|^2     \label{eq:99}
\eeq
Using the form factors $F_0^{BK^*}$ and $A_0^{BK^*}$ given in Eqs.
(\ref{eq:96})
and (\ref{eq:97}), we compute the ratio $R_{\eta_c}$ by varying the parameters
$\Lambda_1$ and $\Lambda_2$ inside the allowed domains discussed in section V.
For the scenario $n_2 =2$, the bounds on $R_{\eta_c}$ are
\beq
1.02 \leq \hs{2mm} R_{\eta_c} \hs{2mm} \leq 2.57 \label{eq:100}
\eeq
For the other scenarios $n_2 = 1$ and $n_2 = 0$, the bounds on $R_{\eta_c}$ are
contained inside the inequality  Eq.(\ref{eq:100}).
It turns out that the ratio $R_{\eta_c}$ being only weakly scenario dependent,
hence the bounds
Eq.(\ref{eq:100}) remain valid for all cases.

\vs{2mm}
$5^o)$
The comparison of the $K(K^*) + \eta_c$ and $K(K^*) + J/\Psi$ decay modes
depends on
the ratio of the decay constants $f_{\eta_c}$ and $f_{J/\Psi}$.
Unfortunately $f_{\eta_c}$ is not experimentally known
and we use theoretical estimates if we want to make predictions.
Using quark model considerations  \cite{REF17} we take
\beq
\frac{f_{\eta_c}}{f_{J/\Psi}} = 0.99 \label{eq:101}
\eeq
Consider first the ratio $T$ defined by :
\beq
T = \frac{\Gamma(B \ra K + \eta_c)}{\Gamma(B \ra K + J/\Psi)} \label{eq:102}
\eeq
Using Eqs.(\ref{eq:5}) and (\ref{eq:6}), we obtain
\beq
T = 2.52 \hs{2mm} \left( \frac{f_{\eta_c}}{f_{J/\Psi}} \right)^2 \hs{2mm}
\left| \frac{F_0^{BK}(m_{\eta_c}^2)}{F_1^{BK}(m_{J/\Psi}^2)} \right|^2
\label{eq:103}
\eeq
In our model $F_0^{BK}$ is constant and $F_1^{BK}$ has a monopole $q^2$
dependence
with the pole mass $\Lambda_F$.

As a consequence we simply have
\beq
\frac{F_0^{BK}(m_{\eta_c}^2)}{F_1^{BK}(m_{J/\Psi}^2)}
= 1 - \frac{m_{J/\Psi}^2}{\Lambda_F^2} \             \label{eq:104}
\eeq
Using the estimate Eq.(\ref{eq:101}),
 we obtain the following bounds of $T$ for the scenario $n_2 = 2$
($5 \hs{1mm} GeV \leq \Lambda_F \leq 5.71 \hs{2mm} GeV$)
\beq
0.94 \hs{2mm} \leq \hs{2mm} T \hs{2mm} \leq \hs{2mm} 1.24 \label{eq:105}
\eeq
 For the scenarios $n_2 = 1$ and $n_2 = 0$, the bounds of $T$ satisfy
the double inequality Eq.(\ref{eq:105}).
Conversely a measurement of the ratio $T$ may provide an opportunity to
extract,
from experiment, the scalar decay constant $f_{\eta_c}$.

Finally we introduce a third ratio $T^*$ defined by
\beq
T^* = \frac{\Gamma(B \ra K^* + \eta_c)}{\Gamma(B \ra K^* + J/\Psi)}
\label{eq:106}
\eeq
Using Eqs.(\ref{eq:8}) and (\ref{eq:9}), we get
\beq
T^* = 0.87 \hs{2mm}
\left( \frac{f_{\eta_c}}{f_{J/\Psi}}  \right)^2
\left| \frac{A_0^{BK^*}(m_{\eta_c}^2)}{A_1^{BK^*}(m_{J/\Psi}^2)} \right|^2
\hs{2mm}
\frac{1}
{(a - b x)^2 + 2 ( 1 + c^2 y^2)} \ \label{eq:107}
\eeq
The ratio $T^*$ depends on the three parameters
$\Lambda_1$, $\Lambda_2$ and $\Lambda_V$.
Varying these quantities inside the allowed domains discussed in section V,
we can obtain bounds for $T^*$.
The result is moderately scenario-dependent and using the estimate
Eq.(\ref{eq:101}),
we obtain
\beq
0.45 \hs{2mm} \leq \hs{2mm} T^* \hs{2mm} \leq \hs{2mm} 0.86 \label{eq:108}
\eeq

\vs{2mm}
$6^o)$
The ratios $R_{\eta_c}$, $T$ and $T^*$ have been discussed by us in a recent
paper
\cite{GKP1}
in order to propose a test of factorization.
However in our previous calculations \cite{GKP1},
the ranges of values for the scenario dependent factors
(denoted there as $S_V$ and $S_A$)
have been underestimated and our previous predictions\cite{GKP1}
for the ratios
$R_{\eta_c}$, $T$ and $T^*$ are different from those obtained here.
For details see Ref.\cite{InterReport}.

\vs{5mm}
\large{}
{\bf  VIII. \hs{2mm} $D \ra \ol{K}(\ol{K}^{*})$ hadronic form factors
and semi-leptonic decays }
\vs{0.5cm}
\normalsize

\vs{2mm}
$1^o)$
The $B \ra K(K^*)$ and $D \ra K(K^*)$ hadronic form factors are related by the
SU(2)
heavy flavor symmetry of Isgur-Wise.
{}From the considerations of section III, it is clear that the $q^2$ dependence
for the form factors $F_1, A_1, A_2$ and $V$ is the same in the $B$ and $D$
sectors :
same values for $n_F, n_1, n_2$ and $n_V$,
pole masses are related by Eqs.(\ref{eq:42})
and (\ref{eq:49}).

These new features of $q^2$ dependences in the $B$ sector obtained so far,
can now be used backwards for analysing the semi-leptonic decays
$D \ra \ol{K} + \ell^{+} + \nu_{\ell}$ and $D \ra \ol{K}^{*} + \ell^{+} +
\nu_{\ell}$.
Using the dimensionless variable $t = q^2/m_D^2$, we introduce the normalized
$q^2$
distribution $X(t)$
\beq
X(t) = \frac{1}{\Gamma_{\it sl}} \hs{2mm}
\frac{d}{dt} \hs{2mm} \Gamma_{\it sl}       \label{eq:109}
\eeq
$X(t)$  is independent, in particular,
on the Cabibbo-Kobayashi-Maskawa parameters
and on the normalizations $F_1^{DK}(0)$ and $A_1^{DK^*}(0)$.

We recall that the quantities $x^D(0)$, $y^D(0)$ and
$z^D(0)$
(used in this paper for normalizing the $B$ sector)
 have been extracted from
experimental data on semi-leptonic decay in the $D$ sector
in a scenario-dependent way,
 because the variation with $q^2$ of the form factors
$F_1^{DK}$, $A_1^{DK^*}$, $A_2^{DK^*}$ and $V^{DK^*}$
has not been measured.

\vs{2mm}
$2^o)$
The $q^2$ distribution for the semi-leptonic decay
$D \ra \ol{K} + \ell^{+} + \nu_{\ell}$
depends on the hadronic form factor $F_1^{DK}(q^2)$.
Defining the dimensionless parameters
\beq
r = \frac{m_K}{m_D} \hs{30mm} \alpha_F = \frac{m_D^2}{\Lambda_{DF}^2}
\label{eq:110}
\eeq
where $\Lambda_{DF}$ is the pole mass in the $D$ sector related to $\Lambda_F$
in the $B$ sector by equation (\ref{eq:42}),
we obtain for $X(t)$ the expression
\beq
X(t) = \frac{1}{I(\alpha_F)} \ \hs{2mm}
\frac{[(1 + r^2 - t)^2 - 4 r^2]^{3/2}}{(1 - \alpha_F \hs{1mm} t)^2} \
\label{eq:111}
\eeq
where we have used, for $F_1^{DK}(q^2)$, a monopole $q^2$ dependence, $n_F =
1$.
The integral $I(\alpha_F)$ is defined by the normalization condition $X(t)$ :
\beq
I(\alpha_F) = \int^{(1 - r)^2}_{0} \hs{2mm}
\frac{[(1 + r^2 - x)^2 - 4 \hs{1mm} r^2]^{3/2}}{(1 - \alpha_F \hs{1mm} x)^2}
\hs{2mm} dx              \label{eq:112}
\eeq
$I(\alpha_F)$ can be computed analytically \cite{InterReport} or numerically.

The distribution $X(t)$ for the semi-leptonic decay
$D \ra \ol{K} + \ell^{+} + \nu_{\ell}$
is represented in Figure 7 for  values of $\alpha_F$
corresponding to the bounds on $\Lambda_F$ obtained in section V
for the three scenarios $n_2 = 2, 1, 0 $ and represented on
Figures 4, 5 and 6.
The distribution $X(t)$ is a monotonically decreasing function of $t$.
Its shape is not very sensitive to $\Lambda_F$ excepted in the neighbourhood
of $t=0$ ($q^2 = 0$).

An estimate of the slope of the $q^2$ distribution
in $D \ra \ol{K} + \ell^{+} + \nu_{\ell}$ at $q^2 = 0$
has been given by Witherell \cite{Witherell}
using two models for the $q^2$ dependence of $F_1^{DK}(q^2)$.
His pole masses $\Lambda_{DF}$ in the $D$ sector correspond
 in our language to $\Lambda_F$ of the $B$ sector :
\beq
5.02 \hs{2mm} GeV \hs{2mm} \leq \hs{2mm}
\Lambda_F \hs{2mm} \leq \hs{2mm} 5.36 \hs{2mm} GeV \label{eq:113}
\eeq
Our three scenarios are in agreement with his range Eq.(\ref{eq:113}).

\vs{2mm}
$3^o)$
The $q^2$ distribution in the semi-leptonic decay
$D \ra \ol{K}^{*} + \ell^{+} + \nu_{\ell} $
depends on the three hadronic form factors
$A_1^{DK^*}(q^2)$, $A_2^{DK^*}(q^2)$ and $V^{DK^*}(q^2)$.
Of course the final vector meson $K^*$
may have three possible polarization states,
$\lambda = 0, \pm 1$.

As previously we define dimensionless parameters and in particular
\beq
\alpha_j = \frac{m_D^2}{\Lambda_{Dj}^2} \hs{20mm} j = 1, 2, V \label{eq:114}
\eeq
where the $\Lambda_{Dj}$ are the pole masses in the $D$ sector related to
$\Lambda_j$ in the $B$ sector by Eq.(\ref{eq:49}).
The formalism is similar to the previous case, although more complicated
because of the $K^*$ polarization.
For details see Ref.\cite{InterReport}.

We have computed $X(t)$ for the three scenarios
$[n_1, n_2, n_V] = [-1, 2, 2], [-1, 1, 2] $ and $[-1, 0, 2] $
using the PDG values \cite{RPR} for $x^D(0)$ and $y^D(0)$.
The parameters $\alpha_1$, $\alpha_2$, $\alpha_V$
- or equivalently $\Lambda_1, \Lambda_2, \Lambda_V$ -
are constrainted to be inside the allowed domains represented on
Figures 1, 2 and 3.
The results are shown on Figures 8, 9 and 10 for the three scenarios
$n_2 = 2, 1, 0$ respectively.
As in the previous case the largest sensitivity of $X(t)$ to the parameters
$\alpha_j$ is in the neighbourhood of \hs{1mm}$t = 0$.

In a similar way, we can study the $q^2$ distribution for the polarization
parameters
$\rho_L^{\it sl}(t)$ and $\cA_{LR}^{\it sl}(t)$.
We only give here the corresponding integrated ones and the results
for the three scenarios $n_2 = 2, 1, 0$ are the following :
\beqa
\hs{3mm}(i) \hs{2mm} n_2 = 2 : \hs{10mm}
& &
0.516 \hs{1mm} \leq \hs{1mm} \rho_L^{\it sl} \hs{1mm} \leq \hs{1mm} 0.541
                                          \label{eq:115} \\
& &
0.885 \hs{1mm} \geq \hs{1mm} \cA_{LR}^{\it sl} \hs{1mm} \geq \hs{1mm} 0.829 \no
\\
\cr
\hs{2mm}(ii) \hs{2mm} n_2 = 1 : \hs{10mm}
& &
0.526 \hs{1mm} \leq \hs{1mm} \rho_L^{\it sl} \hs{1mm} \leq \hs{1mm} 0.541
                                         \label{eq:116} \\
& &
0.904 \hs{1mm} \geq \hs{1mm} \cA_{LR}^{\it sl} \hs{1mm} \geq \hs{1mm} 0.857 \no
\\
\cr
\hs{1mm}(iii) \hs{2mm} n_2 = 0 : \hs{10mm}
& &
0.536 \hs{1mm} \leq \hs{1mm} \rho_L^{\it sl} \hs{1mm} \leq \hs{1mm} 0.538
                                       \label{eq:117} \\
& &
0.904 \hs{1mm} \geq \hs{1mm} \cA_{LR}^{\it sl} \hs{1mm} \geq \hs{1mm} 0.892 \no
\eeqa
In Eqs.(\ref{eq:115}) - (\ref{eq:117}) the results are presented in such a way
to
exhibit a correlation between the largest (smallest) $\rho_L^{\it sl}$
and the smallest (largest) $\cA_{LR}^{\it sl}$.

We observe that the results are moderately scenario dependent
for the three considered cases
and all together we obtain :
\beqa
& & 0.52 \hs{1mm} \leq \hs{1mm} \rho_L^{\it sl} \hs{1mm} \leq \hs{1mm} 0.54 \no
\\
& &                                             \label{eq:118} \\
& & 0.90 \hs{1mm} \geq \hs{1mm} \cA_{LR}^{\it sl} \hs{1mm} \geq \hs{1mm} 0.83
\no
\eeqa

\vs{5mm}
\large{}
{\bf IX. \hs{2mm} Concluding remarks }
\vs{0.5cm}
\normalsize

\vs{2mm}
$1^o)$
Let us first summarize the assumptions and constraints contained in our model.

\vs{2mm}
\hs{2mm} (A) Assumptions :

(a). \hs{1mm}
Factorization holds for color supressed $B$ decays and final state strong
interaction effects can be neglected.

(b). \hs{1mm}
The SU(2) heavy flavor symmetry between the $b$ and $c$ quarks is realized
by the Isgur-Wise relations \cite{IW}.

(c). \hs{1mm}
The normalizations of the hadronic form factors in the $B$ and $D$ sector
are determined by their values at $q^2 = 0$ in the $D$ sector from
 semi-leptonic decays
$D \ra \ol{K} + \ell^{+} + \nu_{\ell}$ and
$D \ra \ol{K}^{*} + \ell^{+} + \nu_{\ell}$.

\vs{2mm}
\hs{2mm} (B) The experimental constraints are :

(d). \hs{1mm}
The experimental rates for $B \ra K + J/\Psi$, $B \ra K^* + J/\Psi$,
$B \ra K + \Psi^{'}$ and $B \ra K^{*} + \Psi^{'}$ used in the form
of ratios of rates $R_{J/\Psi}$, $R_{\Psi^{'}}$, $S$ and $S^*$.

(e). \hs{1mm}
The observed longitudinal polarization fraction $\rho_L$ in $B \ra K^* +
J/\Psi$.

\vs{2mm}
\hs{2mm} (C) The theoretical constraints are :

(f). \hs{1mm}
The explicit form of the $q^2$ dependence of the hadronic form factors $F_1$,
$A_1$, $A_2$, $V$ choosen as $\left[ 1 - q^2/\Lambda^2 \right]^{-n}$ with
$ n = -1, 0, 1, 2$.

(g). \hs{1mm}
The pole masses $\Lambda$ of the various form factors in the $B$ sector are
limited to
the $(5 - 6) \hs{2mm} GeV$ range in order to relate them in a likely way to
$b\ol{s}$
bound state masses.

(h). \hs{1mm}
The ratios of form factors $\mu^B(q^2)$ and $\lambda^B(q^2)$ defined in
Eqs.(\ref{eq:37})
and (\ref{eq:52}) are assumed to be independent of $q^2$ and related in a
natural way
to the pole masses $\Lambda_F$ and $\Lambda_2$ by Eqs.(\ref{eq:44}) and
(\ref{eq:56}).

\vs{2mm}
$2^o)$
Let us make some comments concerning the assumption (c).
The correct procedure for a given scenario in the $B$ sector is to determine
the ratios $x^D(0)$, $y^D(0)$ and $z^D(0)$ by using the same scenario
in the $D$ sector for the analysis of experimental data of semi-leptonic
decays $D \ra \ol{K}(\ol{K}^{*}) +\ell^{+} + \nu_{\ell}$.
Unfortunately we were not able to follow this line
because the only available information on $x^D(0)$, $y^D(0)$ and $z^D(0)$
given in Table 3 has been obtained from experiments, assuming a monopole $q^2$
dependence for all the form factors.
We know however that such a scenario $[n_1, n_2, n_V] = [1, 1, 1]$
is in contradiction with experiment in the $B$ sector.
As other authors \cite{Cheng}, we have used the values of Table 3.
However some theoretical uncertainty on these ratios has to be added to
errors given in Table 3 but it is not an easy task to estimate
such an uncertainty.
We refer to our report \cite{InterReport} for some comments on the
determination of
$F_1^{DK}(0)$ and $A_1^{DK^*}(0)$ from the experimental integrated rates.
These determinations depend moderately in fact on scenarios.

\vs{2mm}
$3^o)$
The $q^2$ dependence of the hadronic form factors in the $D$ sector,
$F_1^{DK}(q^2)$, $A_1^{DK^*}(q^2)$, $A_2^{DK^*}(q^2)$ and $V^{DK^*}(q^2)$
could be
in principle determined, from experiment, by the measurement of all possible
$q^2$ distributions in the semi-leptonic decay,
$D \ra \ol{K} + \ell^{+} + \nu_{\ell}$ ( for $F_1^{DK}$ )
and $D \ra \ol{K}^{*} + \ell^{+} + \nu_{\ell}$ where the final vector meson
is polarized ( for $A_1^{DK^*}$, $A_2^{DK^*}$ and $V^{DK^*}$ ).
The knowledge of the $q^2$ dependence of the hadronic form factors in the $D$
sector,
even with unavoidable errors will help in making the selection of scenarios
in the $D$ sector and, by SU(2) heavy flavor symmetry, in the $B$ sector.

To our knowledge, such an experimental information is not available.
It would be of considerable help for clarifying the theoretical constraints (f)
and (g).

\vs{2mm}
$4^o)$
Let us point out that the various ratios studied in this paper
have different types of dependence
with respect to the form factor values at $q^2 = 0$ in the $D$ sector.

\hs{5mm} (i). \hs{3mm}
$\cA_{LR}$ and $\cA_{LR}^{'}$ depend only on $y^D(0)$.

\hs{4mm} (ii). \hs{3mm}
$\rho_{L}$ and $\rho_{L}^{'}$ depend on $x^D(0)$ and $y^D(0)$.

\hs{3mm} (iii). \hs{3mm}
$S^*$ and $T^*$ depend on $x^D(0)$ and $y^D(0)$.

\hs{4mm} (iv). \hs{3mm}
$R_{\eta_c}$ depends on $x^D(0)$ and $z^D(0)$.

\hs{5mm} (v). \hs{3mm}
$R_{J/\Psi}$ and $R_{\Psi^{'}}$ depend on $x^D(0)$, $y^D(0)$ and $z^D(0)$.

\hs{4mm} (vi). \hs{3mm}
$S$ and $T$ are independent of these three ratios.

For the semi-leptonic normalized distribution $X(t)$, it is independent
on these ratios in the
$D \ra \ol{K} + \ell^{+} + \nu_{\ell}$ case and it depends on $x^D(0)$ and
$y^D(0)$
in the $D \ra \ol{K}^{*} + \ell^{+} + \nu_{\ell}$ mode.

\vs{2mm}
$5^o)$
Among the 64 possible cases, we finally obtain three surviving scenarios
$[n_F, n_1, n_2, n_V] = [+1, -1, n_2, +2]$
corresponding to $n_2 = +2, +1, 0$.
The quality of the fit is very good for $n_2 = 2$, acceptable for $n_1 = 1$ and
marginal for $n_2 = 0$
as illustrated in Figures $1 - 6$.
An improvement of the rate measurements, in particular those involving the
$\Psi^{'}$
, may imply important restrictions for the $\Lambda_1$, $\Lambda_2$,
$\Lambda_V$
and $\Lambda_F$, parameter space and possibly eliminate some scenarios starting
with
$n_2 = 0$.
It is clear that measurements of the quantities
 $\cA_{LR}$, $\cA_{LR}^{'}$, $\rho_L^{'}$
and $R_{\eta_c}$ when compared to those predicted by our model
would considerably help
in reducing the size of the allowed domains
in the $\Lambda_{j}$ parameter space.

The best situation would be to select at the end only one scenario with a small
non empty domain in the $\Lambda_{j}$ parameter space.
The worse situation for our model would be that these new measurements
$\cA_{LR}$, $\cA_{LR}^{'}$, $\rho_L^{'}$ and $R_{\eta_c}$
exclude the three presently
remaining scenarios.

Our model is certainly not the unique way to
analyse the $B \ra K(K^*)$ hadronic form factors.
However if we are in the best situation previously mentioned, it will be
necessary to provide a theoretical support to
the so determined hadronic form factors
and for that, results of Ref.\cite{QCDSUM} seem to be
in a good shape because of the unusual $q^2$ behaviour prediction for
$A_1(q^2)$.

If we are in the worse situation, it will be reasonable to think seriously
of the role played by non-factorizable contributions.

%
%
%
%
\vspace{1cm}
\hspace{1cm} \Large{} {\bf Acknowledgements}    \vspace{0.5cm}

\normalsize{

Y. Y. K would like to thank the Commissariat \`a l'Energie Atomique
of France for the award of a fellowship
and especially  G. Cohen-Tannoudji and J. Ha\"{\i}ssinski
for their encouragements.
}

\newpage
%

\newpage
\section*{Figure captions}
\normalsize
\vspace{0.5cm}

\ben

\item
{\bf Figure 1} :
The allowed domain in the $\Lambda_1$, $\Lambda_2$, $\Lambda_V$ space
due to the constraints $\rho_L + \sam \rho_L \geq 0.70$
and $R_{J/\Psi} - \sam R_{J/\Psi} \leq 2.20$ for $\Lambda_2, \Lambda_V$ $\in$
(5 - 6) $GeV$, $\Lambda_1 \geq 5 \hs{2mm} GeV$.
The scenario is $[n_1, n_2, n_V] = [-1, 2, 2]$.
$\sam \rho_L$ and $\sam R_{J/\Psi}$ are theoretical errors
induced by experimental errors of $x^D(0), y^D(0)$ and $z^D(0)$.

\item
{\bf Figure 2} :
Same as Figure 1 for the scenario [-1, 1, 2].

\item
{\bf Figure 3} :
Same as Figure 1 for the scenario [-1, 0, 2].

\item
{\bf Figure 4} :
The allowed domain in the $\Lambda_F$, $\Lambda_2$, $\Lambda_V$ space
due to the constraint \\
$\Lambda_{1, \hs{1mm}MIN}(\Lambda_2, \Lambda_V, \Lambda_F = 5 \hs{2mm} GeV)
\leq \Lambda_1 \leq \Lambda_{1, \hs{1mm}MAX}(\Lambda_2, \Lambda_V)$
with $\Lambda_F \geq 5 \hs{1mm} GeV$. \\
The scenario is $[n_1, n_2, n_V] = [-1, 2, 2]$.

\item
{\bf Figure 5} :
Same as Figure 4 for the scenario [-1, 1, 2].

\item
{\bf Figure 6} :
Same as Figure 4 for the scenario [-1, 0, 2].

\item
{\bf Figure 7} :
The normalized dimensionless distribution $X(t)$ for the semi-leptonic
decay $D \ra \ol{K} + \ell^{+} + \nu_{\ell}$.
The scenario $n_2 = 2$ corresponds to
$5 \hs{2mm} GeV \hs{1mm} \leq \hs{1mm} \Lambda_F \hs{1mm} \leq \hs{1mm}
5.71 \hs{2mm} GeV$, the scenarios $n_2 =1 $
to
$5 \hs{2mm} GeV \hs{1mm} \leq \hs{1mm} \Lambda_F \hs{1mm} \leq \hs{1mm}
5.39 \hs{2mm} GeV$ and the scenario $n_2 = 0$ to
$5 \hs{2mm} GeV \hs{1mm} \leq \hs{1mm} \Lambda_F \hs{1mm} \leq \hs{1mm}
5.10 \hs{2mm} GeV$.
By Eq.(\ref{eq:42}) the pole masses $\Lambda_{DF}$ in the $D$ sector
can be obtained from $\Lambda_F$ given here.

\item
{\bf Figure 8} :
The normalized dimensionless distribution
$X(t)$ for the semi-leptonic decay
$D \ra \ol{K}^{*} + \ell^{+} + \nu_{\ell}$.
The scenario is $[n_1, n_2, n_V] = [-1, 2, 2]$.
The thickness of the curve is due to the $\Lambda_1,
\Lambda_2, \Lambda_V $ ranges.
\item
{\bf Figure 9} :
Same as Figure 8 for the scenario [-1, 1, 2].

\item
{\bf Figure 10} :
Same as Figure 8 for the scenario [-1, 0, 2].
\een

\newpage
\section*{Table captions}
\normalsize
\vspace{0.5cm}

\begin{enumerate}

\item
{\bf Table 1} : \\
Experimental data as averaged by PDG \cite{RPR}.

\item
{\bf Table 2} : \\
Experimental data for the ratios $R_{\Psi^{'}}$, $R_{J/\Psi}$, $S$ and $S^*$.

\item
{\bf Table 3} : \\
Input data for the charm sector \cite{RPR}.

\item
{\bf Table 4} : \\
Results of the fit at the extreme values of $\Lambda_1$.

\end{enumerate}

\end{document}